\documentclass{article}

\PassOptionsToPackage{numbers,compress}{natbib}
\usepackage[preprint]{neurips_2026}

\usepackage{bbm}
\usepackage[utf8]{inputenc}
\usepackage[T1]{fontenc}
\usepackage{hyperref}
\usepackage{url}
\usepackage{booktabs}
\usepackage{amsfonts}
\usepackage{amsmath,amssymb}
\usepackage{mathtools}
\usepackage{nicefrac}
\usepackage{microtype}
\usepackage{xcolor}
\usepackage{graphicx}
\graphicspath{{./images/}{./}}
\usepackage{caption}
\usepackage{subcaption}
\usepackage{multirow}
\usepackage{algorithm}
\usepackage{algorithmic}
\usepackage{float}
\usepackage{placeins}
\usepackage{enumitem}

\usepackage{tikz}
\usetikzlibrary{arrows.meta,positioning,shapes.geometric,fit,calc,
  decorations.pathmorphing,backgrounds,shadows.blur,patterns,
  decorations.pathreplacing}

\usepackage[textsize=footnotesize]{todonotes}
\usepackage[most]{tcolorbox}
\usepackage{xspace}

\newif\iffinal
\iffinal
    \newcommand{\XL}[1]{}
    \newcommand{\XLinline}[1]{}

\else
    \newcommand{\XL}[1]{\todo[fancyline,color=green!40]{XL: #1}\xspace}
    \newcommand{\XLinline}[1]{\textcolor{NavyBlue}{[XL: #1]}}
\fi

\title{Protein Thoughts: Interpretable Reasoning with Tree of Thoughts and Embedding-Space Flow Matching for Protein--Protein Interaction Discovery}

\author{%
  Kingsley Yeon \\
  Department of Statistics and CCAM\\
  University of Chicago\\
  Chicago, IL 60637 \\
  \texttt{yeon@uchicago.edu} \\
  \And
  Xuefeng Liu \\
  School of Medicine\\
  Stanford University\\
  Stanford, CA 94305\\
  \texttt{xfl@stanford.edu} \\
  \And
  Promit Ghosal \\
  Department of Statistics\\
  University of Chicago\\
  Chicago, IL 60637 \\
  \texttt{promit@uchicago.edu} \\
}

\begin{document}
\maketitle

\begin{abstract}
Protein–protein interactions (PPIs) govern nearly all cellular process, yet computational methods for identifying binding partners typically produce ranked predictions without mechanistic justification. This creates a fundamental barrier to adoption because biologists cannot assess whether predictions reflect genuine biochemical insight or spurious correlations. We present \textbf{Protein Thoughts}, a framework that reformulates PPI discovery as an interpretable search problem with explicit reasoning. The system decomposes binding evidence into four biologically meaningful signals: sequence similarity reflecting evolutionary relationships, structural complementarity capturing geometric fit, interface balance measuring, and chemical compatibility encoding residue-level interactions. Rather than collapsing these signals into an opaque score, we preserve their individual contributions through a transparent value function that enables both ranking and auditing.
To navigate large candidate spaces efficiently, we introduce hypothesis-guided entropy-regularized Tree-of-Thoughts search. A fine-tuned language model generates search directives from embedding-derived features, classifying candidates as high-priority, exploratory, or skippable. These directives condition a Boltzmann policy that balances exploitation with entropy-driven exploration, while hypothesis-aware pruning prevents premature abandonment of promising candidates. For candidates exhibiting score disagreement, hypothesis-conditioned embedding-space flow matching transports protein embeddings toward the binder manifold, with trajectory statistics combining with hypothesis consistency to produce calibrated scores without expensive structure-based docking.
On the SHS148k benchmark, Protein Thoughts achieves mean best-binder rank of 11.2 versus 47.7 for an entropic tree search baseline (i.e., without thought guidance), a 76\% improvement and for binding prediction the trained value function achieves $91.08 \pm 0.19$ Micro-F1, outperforming existing PPI methods on the same dataset.


\end{abstract}

 \section{Introduction}
Proteins do not act alone. From the transient encounters that propagate signaling cascades to the stable assemblies that form molecular machines, protein--protein interactions orchestrate virtually every cellular process \citep{luck2020,keskin2016}. The human interactome encompasses hundreds of thousands of interactions, and perturbations to this network underlie diseases ranging from cancer to neurodegeneration \citep{vidal2011,sahni2015}. Understanding which proteins interact and why is therefore central to both basic biology and therapeutic development.
Experimental mapping of PPIs remains expensive and time-consuming. Yeast two-hybrid screens, affinity purification mass spectrometry, and proximity labeling methods have collectively identified tens of thousands of interactions, yet coverage remains incomplete \citep{rolland2014,huttlin2021}. The emergence of protein language models \citep{lin2023,rives2021} and structure prediction methods \citep{jumper2021,evans2022} has made computational screening increasingly attractive. The bottleneck is no longer whether we can produce a score--it is whether that score means anything.

\textbf{The interpretability gap.}
Current ML methods for PPI prediction predominantly function as black boxes \citep{sledzieski2021,chen2019}. Even AlphaFold Multimer \citep{evans2022} provides confidence metrics like ipTM and pTM without mechanistic rationale. When a model predicts 85\% binding probability, a biologist cannot assess whether this reflects genuine biochemical insight--complementary binding surfaces, conserved motifs, favorable electrostatics--or spurious correlations. The problem is compounded by interaction heterogeneity: barnase--barstar achieves subpicomolar
affinity through complementary interfaces \citep{buckle1994}, antibody--antigen recognition relies on localized epitope--paratope contacts with minimal global structural similarity \citep{wilson1994}, and disordered regions fold upon binding \citep{wright2015}. A single score cannot capture this diversity. This raises a central question which
we address in this paper:

\emph{Can we build PPI prediction systems that not only rank candidates accurately but also expose the biological reasoning behind each prediction, while exploring conformational flexibility through hypothesis-driven structural perturbations such that domain experts can interrogate and critique?}

\textbf{From ranking to reasoning.}
We propose treating PPI discovery as a search problem with explicit reasoning rather than a ranking problem. In experimental biology, scientists examine evidence, weigh competing hypotheses, revise interpretations, and make decisions informed by mechanistic understanding. Computational tools should support this workflow. The Tree of Thoughts framework \citep{yao2023,long2023} enables exploration of alternative solution paths while maintaining multiple hypotheses and backtracking when needed. Chain-of-thought prompting \citep{wei2022} elicits step-by-step reasoning that can be inspected and critiqued. We adapt these ideas to the protein domain where the thoughts are biological hypotheses--shared evolutionary history, geometric complementarity, or favorable chemistry--and the search explores which hypotheses best explain the evidence. Concurrent work combines tree search with diffusion models in the protein domain \citep{liu2025mctd}, using MCTS to guide masked diffusion for sequence recovery in inverse folding. Our approach differs in two key respects: we target PPI discovery rather than sequence design, and our value function is auditable--decomposing binding evidence into biologically meaningful signals rather than relying on an opaque neural network.


\textbf{Hypothesis conditioned embedding space flow.}
Protein language model embeddings encode rich evolutionary and structural information, yet how binding compatibility manifests geometrically in these high-dimensional spaces remains an open question. Traditional approaches treat embeddings as fixed feature vectors, discarding the manifold structure that distinguishes binders from non-interacting pairs. Two proteins with low static cosine similarity may nonetheless bind strongly through representational compatibility that scalar distance metrics fail to capture. We introduce hypothesis-conditioned embedding-space flow matching as a discriminative alternative. A finetuned language model generates natural language hypotheses about binding potential from interpretable embedding-derived features including similarity metrics, difference statistics, and interaction signals. These hypotheses condition velocity fields that transport protein embeddings along trajectories whose quality reveals binding compatibility. Convergence directives guide embeddings toward the binder manifold; rejection directives produce non-convergent trajectories. Trajectory statistics--final distance to binder centroid, improvement over integration, and convergence rate--provide discriminative scores, creating tight integration between symbolic reasoning and continuous flow-based discrimination.

\textbf{Contributions.}
We present Protein Thoughts (Figure~\ref{fig:overview}), integrating four key innovations:
\begin{itemize}[leftmargin=0.5em]
\item \textbf{Decomposed scoring.} For each protein-protein pair, we compute four interpretable signals (Section~\ref{sec:biological_score}) and preserve their individual contributions throughout the pipeline. An interpretable value function (Section~\ref{sec:value_function}) combines them for ranking while enabling post hoc auditing.
\item \textbf{Entropy regularized tree search.} Rather than greedy ranking, we perform budgeted tree of thought search (Section~\ref{sec:tot_search}) that maintains multiple competing hypotheses, backtracks when branches stagnate, and leaves explicit audit trails of what was explored and why.
 \item \textbf{Hypothesis conditioned embedding space flow.} A finetuned language model generates binding hypotheses that condition velocity fields, exploring embedding-space dynamics through five perturbation strategies, which may indicate conformational compatibility without expensive molecular dynamics (Section~\ref{sec:flow}). 
\item \textbf{Structured explanations.} For high value nodes, the language model generates human readable explanations highlighting supporting evidence and tensions between signals, suitable for biological critique and experimental planning.
\end{itemize}

We validate binding prediction on large-scale PPI benchmarks (Table~\ref{tab:benchmarkresults}), achieving state-of-the-art performance with $\mathbf{85.60 \pm 0.80}$ on SHS27k, $\mathbf{91.08 \pm 0.19}$ on SHS148k, and $\mathbf{98.05 \pm 0.03}$ on STRING, outperforming prior methods including ProLLM.\footnote{We follow the exact experimental setup of ProLLM and prior works in this comparison, using the same dataset splits, evaluation metrics, and reported baselines to ensure a fair and direct comparison.} These results demonstrate strong accuracy and scalability across datasets. We also evaluate mutation effects on SKEMPI v2 and show over $2000\times$ speedup compared to AlphaFold Multimer-based screening.

\section{Related Work}
\label{sec:related}
\label{app:related}

\paragraph{Protein representation learning.}
Studies in protein representation focus on hierarchical structural levels critical for biological function. Some research treats protein sequences as biological language and utilizes transformer architectures to model amino acid interactions \citep{madani2023,notin2022}. Other approaches employ masked language modeling to develop attention mechanisms reflecting protein spatial interaction maps \citep{rives2021,lin2023,vig2020}. Structure oriented methods encapsulate functional attributes and spatial data for tasks like molecule binding \citep{jin2021,kong2022}, protein interface studies \citep{mahbub2022,reau2023}, and property predictions \citep{zhang2022ontoprotein}. ESM-2 \citep{rives2021} and ESMFold \citep{lin2023} provide high quality embeddings that capture evolutionary and structural information. However, most works rely on single modal data, overlooking cross modality interactions among text and protein sequence and structure information.

\paragraph{PPI prediction methods.}
Traditional PPI prediction relied on sequence features \citep{shen2007}, domain interactions \citep{deng2002}, and network properties \citep{sharan2007}. Deep learning methods including D-SCRIPT \citep{sledzieski2021} and Siamese RCNNs \citep{chen2019} learn representations directly from sequences but function as black boxes. Structure based methods like AlphaFold Multimer \citep{evans2022} predict complex geometry but provide confidence metrics without mechanistic rationale. Gromov Wasserstein approaches \citep{peyre2019} offer principled distance measures between proteins but require careful feature engineering to achieve discrimination.

\paragraph{Large language models for proteins.}
Recent advances in LLMs have extended to protein research. ProteinChat \citep{guo2023proteinchat} and ProtChatGPT \citep{wang2024protchatgpt} align protein structure or sequence data with textual descriptions. ProtST \citep{xu2023protst} focuses on sequence based modeling. These methods enable natural language interaction with protein data but typically do not provide interpretable decomposition of binding evidence or explicit reasoning traces. Our work differs by using the finetuned LLM to generate hypotheses that guide conformational exploration rather than simply describing proteins.

\paragraph{Tree search and reasoning.}
Tree of Thoughts \citep{yao2023,long2023} enables explicit exploration of alternative solution paths in language model reasoning. Chain of thought prompting \citep{wei2022} elicits step by step reasoning. Monte Carlo tree search has been combined with diffusion models for protein design \citep{liu2025mctd}. Entropy regularized reinforcement learning \citep{ziebart2010,haarnoja2018} provides principled exploration exploitation tradeoffs. We adapt these ideas to PPI discovery where thoughts are biological hypotheses and the search explores which hypotheses best explain binding evidence.

\paragraph{Flow matching and generative models.}
Flow matching \citep{lipman2023,liu2023flow} provides efficient training of continuous normalizing flows. Diffusion models have been applied to protein structure generation \citep{watson2023} and sequence design \citep{hsu2022}. Our hypothesis conditioned conformational flow differs by conditioning velocity fields on natural language hypotheses generated by a finetuned LLM, creating tight integration between symbolic reasoning and continuous optimization.

\section{Methods}
\label{sec:methods}

\subsection{Decomposed Scoring}\label{sec:biological_score}

\begin{figure}[h]
\centering
\includegraphics[width=\textwidth]{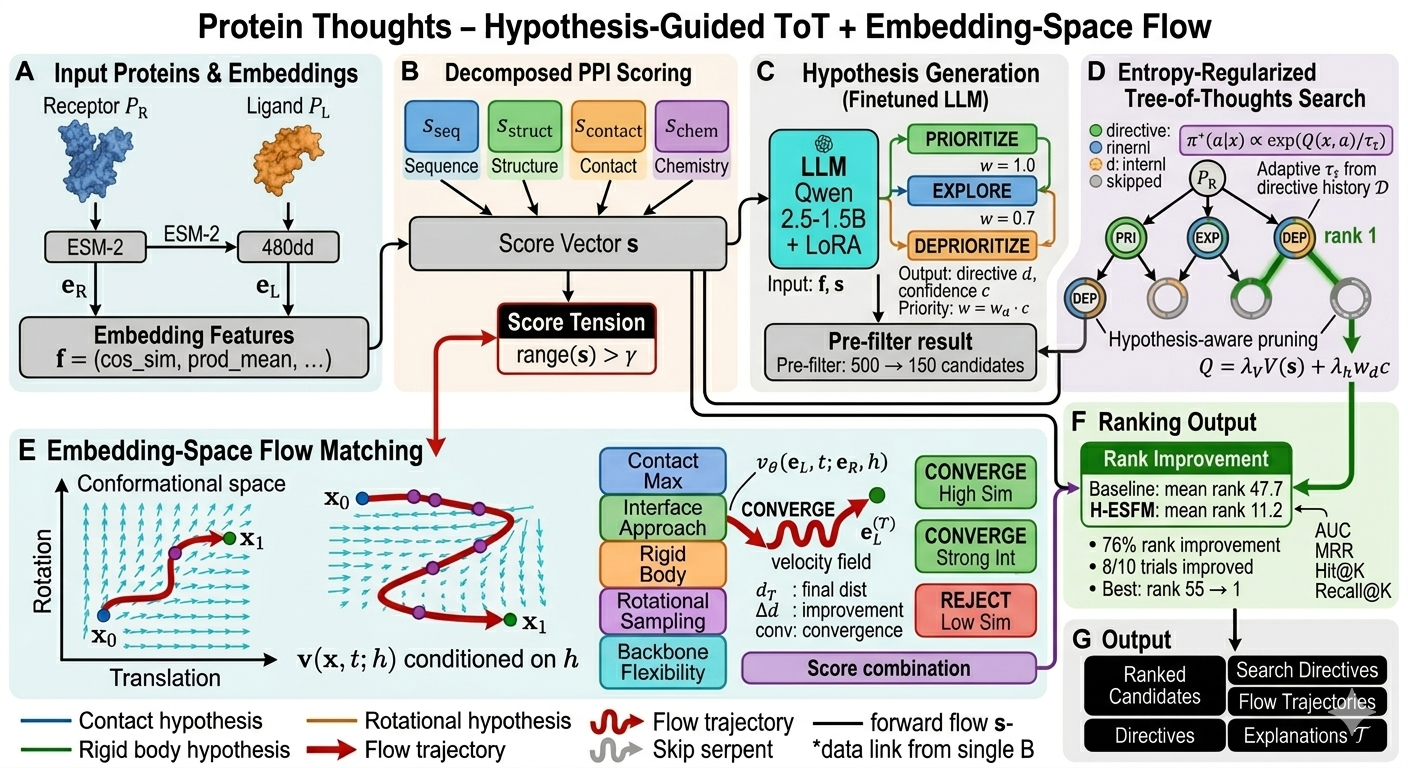}
\caption{\textbf{Protein Thoughts pipeline: hypothesis-guided entropy-regularized Tree-of-Thoughts search with embedding-space flow matching.} (see Algorithm~\ref{alg:tot})
\textbf{A,} Input proteins are encoded via ESM-2 into 480-dimensional embeddings $\mathbf{e}_R, \mathbf{e}_L$, from which interpretable features (cosine similarity, product statistics) are computed.
\textbf{B,} Decomposed PPI scoring computes four interpretable metrics. Score tension (high variance across components) triggers selective flow matching. 
\textbf{C,} A fine-tuned language model (Qwen 2.5-1.5B with LoRA) generates search directives (PRIORITIZE, EXPLORE, DEPRIORITIZE, SKIP) with confidence scores, enabling pre-filtering.
\textbf{D,} Entropy-regularized Tree-of-Thoughts search uses a Boltzmann policy $\pi^*(a|x) \propto \exp(Q/\tau_t)$ with hypothesis-weighted Q-values and adaptive temperature. Hypothesis-aware pruning requires both score stagnation and directive agreement.
\textbf{E,} For candidates with score tension, hypothesis-conditioned embedding-space flow matching transports ligand embeddings toward the binder manifold. The velocity field $v_\theta(\mathbf{e}_L, t; \mathbf{e}_R, h)$ is conditioned on CONVERGE/REJECT hypotheses, with trajectory statistics (final distance, improvement, convergence) combined with hypothesis consistency.
\textbf{F,} On SHS148k, the system achieves mean rank 11.2 versus 47.7 baseline (76\% improvement).
\textbf{G,} Output includes ranked candidates, search directives, flow trajectories, and natural language explanations in audit tree $\mathcal{T}$.}
\label{fig:overview}
\end{figure}

For each protein-protein pair $(P_R, P_L)$ we compute four biologically motivated scores $\mathbf{s} = (s_\mathrm{seq}, s_\mathrm{struct}, s_\mathrm{contact}, s_\mathrm{chem}) \in [0,1]^4$ (detailed in Appendix~\ref{sec:scoring}), each capturing an orthogonal aspect of molecular representation (Figure~\ref{fig:overview}, panel B). These signals were chosen not for maximal predictive power but for interpretability. Each corresponds to a biological concept that practitioners routinely consider and can question using domain expertise.

\textbf{Interface-residue sequence similarity.}
We compute $s_\mathrm{seq}$ via global Needleman--Wunsch alignment \citep{needleman1970} with BLOSUM62, normalized by maximum sequence length. High values reflect coevolutionary history, shared binding motifs, and conserved interaction patterns \citep{marks2011,dejuan2013,cong2019}. Members of the immunoglobulin superfamily share characteristic sequence patterns reflecting common ancestry and similar interaction modes \citep{williams1988}, and in pharmacology, sequence similarity predicts cross-reactivity where a drug inhibiting one kinase may affect related kinases with similar active sites \citep{karaman2008}.

\textbf{Structural complementarity.}
We define $s_\mathrm{struct} = \exp(-\mathrm{RMSD}/\sigma)$ with $\sigma = 5$\,\AA, computed via the Kabsch algorithm \citep{kabsch1976}. This measures geometric lock-and-key fit \citep{fischer1894} without full docking. Enzyme active sites are shaped to accommodate specific substrates and antibody paratopes are sculpted to match epitope topology \citep{mian1991}; imatinib in the ATP-binding site of BCR--ABL \citep{druker2001} exemplifies high affinity through snug geometric fit.

\textbf{Interface balance.}
We compute $s_\mathrm{contact} = \min(c_R, c_L) / \max(c_R, c_L)$ where $c$ counts surface-exposed residues in structured regions (RSA $>$20\%, pLDDT $>$70). Stable heterodimeric complexes typically bury 1200--2000\,\AA$^2$ per partner \citep{locontechothia1999}, and peptide drugs must engage sufficient interface area to achieve meaningful affinity \citep{walensky2004}.

\textbf{Chemical compatibility.}
We compute $s_\mathrm{chem}$ as the mean pairwise entry of a Miyazawa--Jernigan type matrix $M \in \mathbb{R}^{20 \times 20}$ encoding opposite charges (+1.0), hydrogen bonds (+0.7), hydrophobic contacts (+0.8), and like charges ($-0.5$). Protein interfaces are enriched in salt bridges, hydrogen bonds, and hydrophobic clusters \citep{janin1990,bogan1998,chakrabarti2002}, and drug-resistance mutations often alter interface chemistry in ways that disrupt binding \citep{yun2008}.

\textbf{Designed disagreement.}
These four signals capture orthogonal aspects of molecular recognition and are therefore designed to disagree. A pair might show strong evolutionary signal but poor geometry, or excellent shape complementarity but unfavorable chemistry. These disagreements are not noise to be averaged away. They are structure that reveals the biological nature of the interaction. For instance, when barnase--barstar scores high on geometry and interface balance but modest on chemistry, the system should correctly infer shape driven binding \citep{buckle1994}. When an antibody--antigen pair shows near zero structural similarity but high interface balance, the system should recognize the signature of CDR epitope recognition \citep{wilson1994}.

\subsection{Interpretable Value Function}
\label{sec:value_function}

The value function $V_\phi$ estimates binding likelihood while maintaining interpretability through a mandatory four-dimensional bottleneck aligned with biological scores, ensuring predictions remain explainable via decomposed evidence rather than opaque latent features.

\textbf{Biological scores.} For each protein pair $(A, B)$, we compute four interpretable scores $\mathbf{s} = (s_\text{seq}, s_\text{struct}, s_\text{contact}, s_\text{chem}) \in [0,1]^4$ characterizing sequence compatibility, structural alignment, contact propensity, and chemical complementarity respectively.

\paragraph{Architecture (PPIProjectedNet).}
Protein sequences are embedded using ESM-2~\cite{lin2023evolutionary} (35M parameters), producing mean-pooled representations $\mathbf{e}_A, \mathbf{e}_B \in \mathbb{R}^{480}$. The pair feature vector concatenates embeddings with their difference, product, and biological scores:
 $  \mathbf{x} = \bigl[\mathbf{e}_A \;\|\; \mathbf{e}_B \;\|\; \mathbf{e}_A - \mathbf{e}_B$ and $
    \;\|\; \mathbf{e}_A \odot \mathbf{e}_B \;\|\; \mathbf{s}\bigr] \;\in\; \mathbb{R}^{1924}$
where $\|$ denotes concatenation, $\mathbf{e}_A - \mathbf{e}_B$ encodes asymmetric relationships, and $\mathbf{e}_A \odot \mathbf{e}_B$ captures feature-wise interactions. Each embedding is independently projected to $d_\text{model}{=}256$:
    $\mathbf{h}_A^{(0)} = \text{GELU}\!\left(\text{LN}\!\left(W_A\,\mathbf{e}_A
    + \mathbf{b}_A\right)\right),$ and $
    \mathbf{h}_B^{(0)} = \text{GELU}\!\left(\text{LN}\!\left(W_B\,\mathbf{e}_B
    + \mathbf{b}_B\right)\right)$
where $\text{GELU}(x) = x \cdot \Phi(x)$ with $\Phi(\cdot)$ the standard normal CDF. A two-layer, eight-head transformer encoder~\cite{vaswani2017attention} refines each representation to $\mathbf{h}_A, \mathbf{h}_B \in \mathbb{R}^{256}$, then cross-attention captures complementarity:
\begin{equation}
    \mathbf{h}_\text{cross} = \text{LN}\!\left(\mathbf{h}_A
    + \text{MHA}(\underbrace{\mathbf{h}_A}_{\text{query}},\;
    \underbrace{\mathbf{h}_B}_{\text{key}},\;
    \underbrace{\mathbf{h}_B}_{\text{value}})\right) \;\in\; \mathbb{R}^{256}
\end{equation}
The biological scores are projected as $\mathbf{h}_s = W_s\,\mathbf{s} + \mathbf{b}_s \in \mathbb{R}^{256}$ and fused with protein representations by a two-layer MLP with LayerNorm, GELU, and dropout ($p{=}0.20$):
    $\mathbf{h} = \text{MLP}\!\left([\mathbf{h}_\text{cross} \;\|\; \mathbf{h}_B
    \;\|\; \mathbf{h}_s]\right) \;\in\; \mathbb{R}^{256}$
Critically, $\mathbf{h}$ is projected through a \emph{mandatory four-dimensional bottleneck}:
    $\mathbf{z} = W_\text{btn}\,\mathbf{h} + \mathbf{b}_\text{btn} \;\in\; \mathbb{R}^{4},$ and $
    \quad W_\text{btn} \in \mathbb{R}^{4 \times 256},\; \mathbf{b}_\text{btn} \in \mathbb{R}^{4}$
where $\mathbf{z} = (z_\text{seq}, z_\text{struct}, z_\text{contact}, z_\text{chem})$ mirrors the biological scores but is refined by the ESM-2 backbone. The final prediction passes exclusively through $\mathbf{z}$:
    $\hat{y} = \sigma\!\left(W_\text{cls}\,\text{GELU}(\mathbf{z}) + b_\text{cls}\right),$ and $
     W_\text{cls} \in \mathbb{R}^{1 \times 4},\; b_\text{cls} \in \mathbb{R}$
where $\sigma(\cdot)$ is the sigmoid function. This bottleneck forces every prediction through four interpretable dimensions, ensuring improved performance arises from better weighting of biological signals rather than opaque latent structure.

\subsection{Hypothesis-Guided Tree of Thoughts Search}
\label{sec:tot_search}

Given the decomposed scores and value function, a naive approach would simply rank all candidates by $V(\mathbf{s})$ and return the top $k$. This flat ranking discards context because two candidates with similar $V$ might achieve their scores for entirely different reasons, and those reasons matter for experimental follow up. Moreover, exhaustive evaluation of large candidate pools is computationally prohibitive when each scoring requires embedding computation and neural network inference. We instead formulate PPI discovery as a sequential decision process guided by language model hypotheses and perform structured tree search.

\textbf{Hypothesis-guided pre-filtering.}
Before expensive scoring, a fine-tuned language model (Qwen 2.5-1.5B with LoRA) maps embedding-derived features--cosine similarity and four biological scores from Section~\ref{sec:value_function}--to search directives that guide candidate prioritization (Table~\ref{tab:directives}). Each directive carries a learned priority weight $w_d \in [0.1, 1.0]$ and confidence $c \in [0, 1]$, reducing a large candidate pool to a smaller number of high-priority candidates while preserving recall.

\begin{table}[t]
\centering
\caption{Search directives generated by the hypothesis model. Directives are organized by action type (rows) and primary embedding signal (columns), with priority weights $w_d$ indicating search emphasis. $\alpha_{\texttt{P}}, \alpha_{\texttt{D}}$ denote thresholds for high-priority and deprioritized actions determined via cross-validation.}
\label{tab:directives}
\small
\begin{tabular}{@{}lccc@{}}
\toprule
\textbf{Action} & \textbf{Similarity} & \textbf{Interaction} & \textbf{Score Alignment} \\
\midrule
\textsc{Prioritize} ($w_d \geq \alpha_{\texttt{P}}$) & \texttt{HIGH\_SIMILARITY} & \texttt{STRONG\_INTERACTION} & \texttt{SCORE\_ALIGNMENT} \\
\textsc{Explore} ($w_d \in [\alpha_{\texttt{D}},\alpha_{\texttt{P}}]$) & \multicolumn{2}{c}{\texttt{MODERATE\_SIGNAL}} & \texttt{TENSION\_DETECTED} \\
\textsc{Deprioritize} ($w_d \leq \alpha_{\texttt{D}}$) & \texttt{DISTANT} & \texttt{WEAK\_SIGNAL} & -- \\
\textsc{Skip} ($w_d = 0.1$) & \multicolumn{3}{c}{\texttt{INCOMPATIBLE}} \\
\bottomrule
\end{tabular}
\end{table}

\textbf{State representation.}
A search state encodes the receptor protein, the set of evaluated ligands, and the hypothesis context:
\begin{equation}
x_t = (P_R, \mathcal{V}_t, \mathcal{H}_t), \qquad \mathcal{V}_t \subseteq \mathcal{L},
\end{equation}
where $\mathcal{L}$ denotes the ligand pool and
\begin{equation}
\mathcal{H}_t = \{(a, h_a, d_a) : a \in \mathcal{V}_t\}
\end{equation}
stores the hypothesis $h_a$ and directive $d_a$ associated with each evaluated candidate. The search tree $\mathcal{T} = (V, E)$ records decomposed scores, directives, confidence values, and natural-language explanations at every node.

\textbf{Hypothesis-weighted Boltzmann policy.}
Greedy selection risks missing promising alternatives when scores are noisy or when multiple binding modes exist. We therefore employ entropy-regularized action selection~\citep{ziebart2010,haarnoja2018}, augmented with hypothesis-derived priorities:
\begin{equation}
Q(x, a) = \lambda_V \cdot V(\phi(P_R, a)) + \lambda_h \cdot w_{d_a} \cdot c_a,
\end{equation}
where $V(\phi(P_R, a))$ is the value function score from Section~\ref{sec:value_function}, $w_{d_a}$ is the directive priority weight, and $c_a$ is the associated confidence. Mixing coefficients $\lambda_V = 0.7$ and $\lambda_h = 0.3$ are determined via cross-validation.

Actions are sampled according to the Boltzmann policy
\begin{equation}
\pi^*(a | x) =
\frac{\exp(Q(x, a)/\tau_t)}
{\sum_{b \in \mathcal{A}(x)} \exp(Q(x, b)/\tau_t)}.
\label{eq:boltzmann}
\end{equation}

As $\tau_t \to 0$, the policy becomes increasingly greedy; larger $\tau_t$ promotes exploration. Rather than fixing temperature globally, we adapt it according to observed directive patterns. Let
\begin{equation}
\rho_{\text{tension}}
=
\frac{
|\{d \in \mathcal{D}_t :
d = \texttt{EXPLORE\_TENSION\_DETECTED}\}|
}{
|\mathcal{D}_t|
},
\end{equation}
and
\begin{equation}
\rho_{\text{priority}}
=
\frac{
|\{d \in \mathcal{D}_t :
d \in \texttt{PRIORITIZE\_*}\}|
}{
|\mathcal{D}_t|
},
\end{equation}
where $\mathcal{D}_t$ denotes directives encountered up to step $t$. The adaptive temperature schedule is
\begin{equation}
\tau_t =
\begin{cases}
\tau_0 + \delta_{\text{explore}}
& \text{if }
\rho_{\text{tension}} > \theta_{\text{tension}}, \\[4pt]
\max(\tau_{\min},
\tau_0 - \delta_{\text{exploit}})
& \text{if }
\rho_{\text{priority}} > \theta_{\text{priority}}, \\[4pt]
\tau_0
& \text{otherwise},
\end{cases}
\end{equation}
with $\tau_0 = 0.35$, $\tau_{\min}=0.2$,
$\theta_{\text{tension}} = 0.3$,
$\theta_{\text{priority}} = 0.4$,
$\delta_{\text{explore}} = 0.15$,
and $\delta_{\text{exploit}} = 0.1$ (cross-validated).

\textbf{Hypothesis-aware finite difference pruning.}
Computational budgets are finite. We monitor the value trajectory along each branch using finite-difference signals:
\begin{align}
\Delta V_t &= V_t - V_{t-1}, \\
\Delta^2 V_t &= \Delta V_t - \Delta V_{t-1}.
\end{align}
A branch is pruned if
\begin{equation}
\Delta V_t < \varepsilon_1,
\qquad
\Delta^2 V_t \leq \varepsilon_2,
\end{equation}
and the associated hypothesis directive belongs to
\begin{equation}
d_t \in
\{
\texttt{DEPRIORITIZE\_*},
\texttt{SKIP\_*}
\}.
\end{equation}
This prevents excessive exploration of stagnant branches while preserving candidates whose hypotheses indicate unresolved ambiguity.

\textbf{Selective flow triggering.}
The language model additionally predicts whether embedding-space refinement is warranted. Flow matching is triggered only when the hypothesis recommends flow or when score tension exceeds a threshold:
\begin{equation}
\text{use\_flow}(a)
=
\mathbf{1}[h_a = \texttt{FLOW\_RECOMMENDED}]
\lor
\mathbf{1}[
\text{score\_range}(\mathbf{s}_a)
>
\gamma_{\text{tension}}
],
\end{equation}
where
\begin{equation}
\text{score\_range}(\mathbf{s})
=
\max_i s_i - \min_i s_i,
\end{equation}
and $\gamma_{\text{tension}} = 0.35$ is selected via cross-validation. This selective triggering reduces flow calls by approximately 60\% while concentrating refinement on ambiguous candidates.

\begin{figure}[h]
\centering
\includegraphics[width=0.9\textwidth]{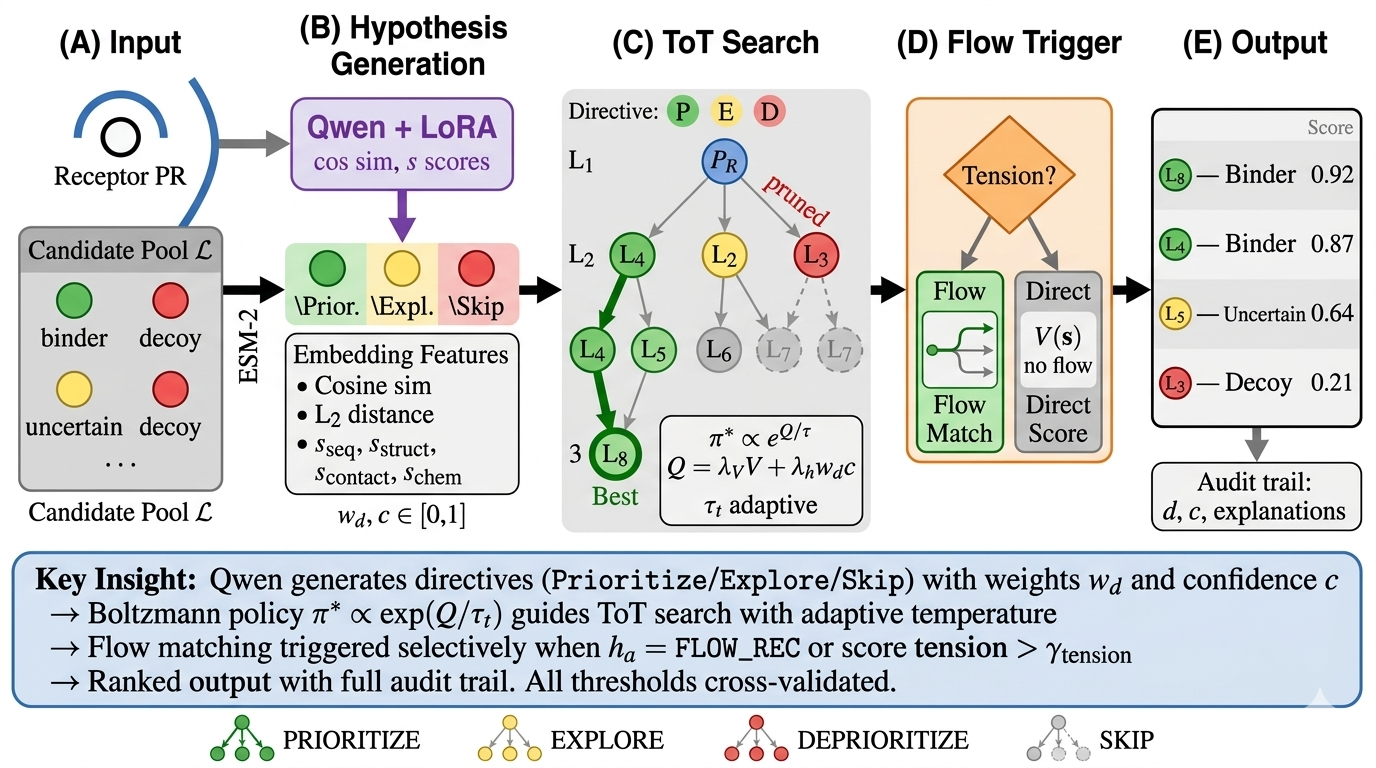}
\caption{Hypothesis-guided entropy-regularized Tree of Thoughts (ToT) search with selective flow matching. 
\textbf{(A)} Input receptor $P_R$ and candidate pool $\mathcal{L}$ are embedded via ESM-2. 
\textbf{(B)} Qwen generates search directives (\textsc{Prioritize}, \textsc{Explore}, \textsc{Deprioritize}, \textsc{Skip}) with priority weights $w_d$ and confidence $c$. 
\textbf{(C)} ToT search expands candidates via Boltzmann policy $\pi^* \propto \exp(Q/\tau_t)$ where $Q = \lambda_V V + \lambda_h w_d c$; adaptive temperature increases for high tension rates and decreases for high priority rates. 
\textbf{(D)} Flow matching triggers selectively when hypotheses or score tensions indicate geometric ambiguity. 
\textbf{(E)} Output: ranked candidates with decomposed evidence and natural-language audit trail.}
\label{fig:search}
\end{figure}

\subsection{Hypothesis-Conditioned Embedding-Space Flow Matching}
\label{sec:flow}

\begin{figure}[h]
\centering
\includegraphics[width=0.9\linewidth]{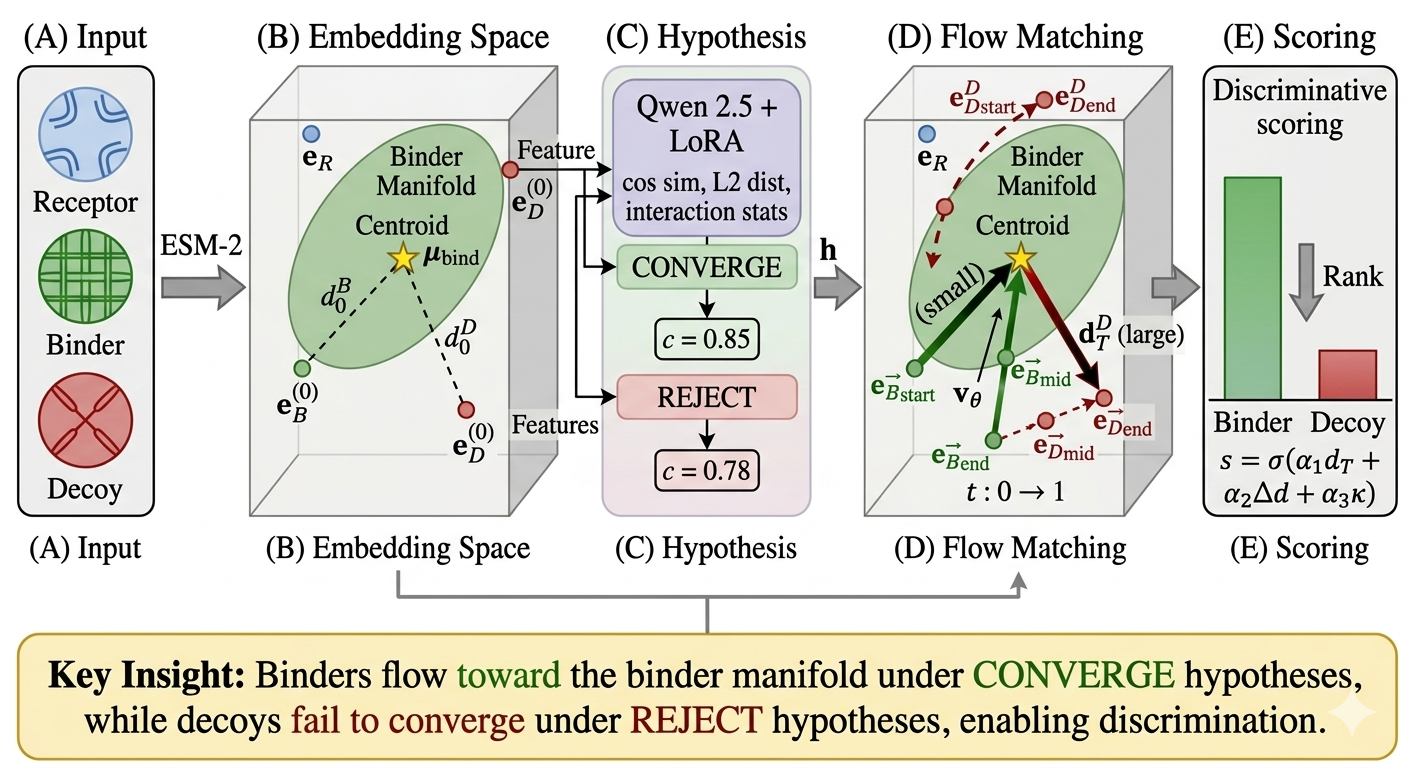}
\caption{Hypothesis-conditioned embedding-space flow matching for PPI discovery. 
\textbf{(A)} Input receptor and candidate proteins. 
\textbf{(B)} ESM-2 embeddings in high-dimensional space with binder manifold centroid $\boldsymbol{\mu}_{\text{bind}}$. 
\textbf{(C)} Qwen generates binding hypotheses from embedding-derived features. 
\textbf{(D)} Flow matching transports embeddings through the learned velocity field $\mathbf{v}_\theta$ conditioned on hypotheses. 
\textbf{(E)} Trajectory statistics and hypothesis consistency jointly produce discriminative scores separating binders from decoys.}
\label{fig:flow_matching}
\end{figure}

Traditional approaches treat protein language model embeddings as fixed feature vectors for downstream prediction. This discards geometric information in the embedding manifold that may encode functional compatibility. In particular, two proteins may have low static cosine similarity or large Euclidean distance while still exhibiting binding-compatible local structure. Conversely, high similarity alone need not imply interaction if the aligned dimensions do not correspond to complementary binding signals. We therefore introduce hypothesis-conditioned embedding-space flow matching, which uses LLM-generated hypotheses to define a directed refinement process in representation space (Figure~\ref{fig:flow_matching}). The resulting trajectories should not be interpreted as literal conformational dynamics; rather, they define learned embedding-space dynamics that may reflect conformational or functional compatibility at the representation level.

\textbf{Embedding feature extraction.}
Each hypothesis type derives from 11 interpretable metrics computed from the receptor-ligand embedding pair $(\mathbf{e}_R, \mathbf{e}_L) \in \mathbb{R}^d \times \mathbb{R}^d$. We compute: \textbf{Similarity metrics.} Cosine similarity $\cos(\mathbf{e}_R, \mathbf{e}_L)$, L2 distance $\|\mathbf{e}_R - \mathbf{e}_L\|_2$, and alignment score measuring mean absolute difference in the top-$k$ most similar dimensions.
\textbf{Difference statistics.} Mean, standard deviation, and maximum absolute value of $\mathbf{e}_R - \mathbf{e}_L$, capturing directional separation patterns.
\textbf{Interaction signal.} Mean, standard deviation, and positive fraction of $\mathbf{e}_R \odot \mathbf{e}_L$, encoding complementary feature activation.
\textbf{Structural indicators.} Norm ratio $\|\mathbf{e}_R\|/\|\mathbf{e}_L\|$ and sparsity (fraction of near-zero product elements).

\textbf{Flow matching formulation.}
We formulate discriminative scoring as a continuous normalizing flow problem~\citep{lipman2023,liu2023flow}, transporting ligand embeddings along learned velocity fields conditioned on binding hypotheses. Following conditional flow matching~\citep{lipman2023}, we define the probability path via optimal transport interpolation:
\begin{equation}
\mathbf{e}_L^{(t)} = (1 - t) \mathbf{e}_L^{(0)} + t \mathbf{e}_L^{(1)}, \quad t \in [0, 1],
\end{equation}
where $\mathbf{e}_L^{(0)}$ is the initial ligand embedding (with optional Gaussian noise $\boldsymbol{\epsilon} \sim \mathcal{N}(0, \sigma^2 I)$ for regularization, $\sigma$ cross-validated) and $\mathbf{e}_L^{(1)}$ is the target embedding. For true binders, $\mathbf{e}_L^{(1)}$ is set to the binder centroid $\boldsymbol{\mu}_{\mathrm{bind}}$ computed as the mean embedding of all confirmed binding partners in the training set; for decoys, $\mathbf{e}_L^{(1)}$ remains the original embedding (i.e., no movement toward the binder manifold is encouraged). This asymmetric target construction encodes the discriminative objective: binders should flow toward the binder manifold while decoys should not.

The hypothesis-conditioned velocity field extends recent molecular flow matching approaches~\citep{watson2023,yim2023se3}:
\begin{equation}
\mathbf{v}_\theta(\mathbf{e}_L, t; \mathbf{e}_R, h) = f_\theta\left( [\mathbf{e}_R; \mathbf{e}_L; \mathbf{e}_R - \mathbf{e}_L; \mathbf{e}_R \odot \mathbf{e}_L; \phi(t); \mathbf{h}] \right),
\end{equation}
where $\mathbf{e}_R$ is the receptor embedding, the difference term $\mathbf{e}_R - \mathbf{e}_L$ encodes directional separation, the element-wise product $\mathbf{e}_R \odot \mathbf{e}_L$ captures feature-wise interactions analogous to attention mechanisms, $\phi(t)$ is a sinusoidal time embedding enabling the network to modulate velocity across the flow trajectory, and $\mathbf{h} \in \mathbb{R}^{d_h}$ is the hypothesis embedding combining directive type, embedding features, and optionally the LLM hidden state ($d_h$ cross-validated).

\textbf{Training objective.} The network $f_\theta$ is trained to predict the conditional vector field $\mathbf{u}_t(\mathbf{e}_L | \mathbf{e}_L^{(1)}) = \mathbf{e}_L^{(1)} - \mathbf{e}_L^{(0)}$ induced by the optimal transport path. The flow matching loss for a single pair is:
\begin{equation}
\mathcal{L}_{\mathrm{FM}} = \mathbb{E}_{t \sim \mathcal{U}(0,1)} \left\| \mathbf{v}_\theta(\mathbf{e}_L^{(t)}, t; \mathbf{e}_R, h) - \mathbf{u}_t \right\|^2.
\end{equation}
We compute this loss separately for positive (binder) and negative (decoy) pairs:
\begin{align}
\mathcal{L}_{\mathrm{FM}}^{(+)} &= \mathbb{E}_{(\mathbf{e}_R, \mathbf{e}_L) \in \mathcal{D}^{+}} \left[ \mathcal{L}_{\mathrm{FM}}(\mathbf{e}_R, \mathbf{e}_L, h^{+}) \right], \\
\mathcal{L}_{\mathrm{FM}}^{(-)} &= \mathbb{E}_{(\mathbf{e}_R, \mathbf{e}_L) \in \mathcal{D}^{-}} \left[ \mathcal{L}_{\mathrm{FM}}(\mathbf{e}_R, \mathbf{e}_L, h^{-}) \right],
\end{align}
where $\mathcal{D}^{+}$ denotes the set of confirmed binder pairs with \textsc{Converge} hypotheses $h^{+}$, and $\mathcal{D}^{-}$ denotes decoy pairs with \textsc{Reject} hypotheses $h^{-}$. The positive loss $\mathcal{L}_{\mathrm{FM}}^{(+)}$ encourages accurate velocity prediction for binders flowing toward the binder manifold, while the negative loss $\mathcal{L}_{\mathrm{FM}}^{(-)}$ trains the model on decoy trajectories that should not converge. To enforce discriminative separation, we augment the reconstruction objective with a margin loss:
\begin{equation}
\mathcal{L} = \mathcal{L}_{\mathrm{FM}}^{(+)} + \lambda_{\mathrm{margin}} \cdot \mathrm{ReLU}\left(\gamma - (\mathcal{L}_{\mathrm{FM}}^{(-)} - \mathcal{L}_{\mathrm{FM}}^{(+)})\right),
\end{equation}
where $\gamma$ is the margin threshold enforcing that decoy reconstruction error exceeds binder reconstruction error by at least $\gamma$, and $\lambda_{\mathrm{margin}}$ balances reconstruction quality against contrastive separation. Intuitively, this loss encourages the velocity field to fit binder trajectories well (low $\mathcal{L}_{\mathrm{FM}}^{(+)}$) while deliberately fitting decoy trajectories poorly (high $\mathcal{L}_{\mathrm{FM}}^{(-)}$), creating a discriminative gap. The margin term activates only when $\mathcal{L}_{\mathrm{FM}}^{(-)} - \mathcal{L}_{\mathrm{FM}}^{(+)} < \gamma$, pushing the model to increase the gap until it exceeds the threshold. Both $\gamma$ and $\lambda_{\mathrm{margin}}$ are determined via cross-validation to balance reconstruction fidelity against discrimination strength.
 
\textbf{Hypothesis embedding architecture.}
The hypothesis conditions the velocity field through a learned embedding that fuses structured predictions with latent representations. Given directive index $d \in \{1, \ldots, 7\}$, feature vector $\mathbf{f} \in \mathbb{R}^{11}$, prediction flags $\mathbf{p} \in \mathbb{R}^4$ (is-binder, is-non-binder, confidence, is-converge), and optional hidden state $\mathbf{h}_{\mathrm{LM}} \in \mathbb{R}^{1536}$, we compute component embeddings each of dimension 48 via learned embedding tables and MLPs, then concatenate and project to the final hypothesis embedding $\mathbf{h} \in \mathbb{R}^{192}$. When hidden states are unavailable (for negatives without cached hypotheses), a separate projection layer operates on the three-component concatenation, maintaining discriminative signal from parsed features while avoiding dimension mismatch.

\textbf{Trajectory integration and scoring.}
We integrate the ODE $\mathrm{d}\mathbf{e}_L/\mathrm{d}t = \mathbf{v}_\theta(\mathbf{e}_L, t; \mathbf{e}_R, h)$ via forward Euler with $T$ steps, tracking the distance trajectory to the binder centroid $\boldsymbol{\mu}_{\mathrm{bind}}$ computed from training positives. Three discriminative statistics are extracted: final distance $d_T$ (proximity to binder manifold), improvement $\Delta d = d_0 - d_T$ (net movement toward binders), and convergence rate $\kappa = \bar{d}_{[0, T/4]} - \bar{d}_{[3T/4, T]}$ (trajectory acceleration). The discriminative score combines these with hypothesis consistency:
\begin{equation}
s = \sigma\left( \alpha_1 \cdot \frac{-d_T}{\tau_d} + \alpha_2 \cdot \frac{\Delta d}{\tau_{\Delta}} + \alpha_3 \cdot \frac{\kappa}{\tau_{\kappa}} \right) \cdot (\beta_0 + \beta_1 \cdot c),
\end{equation}
where $\sigma$ is the sigmoid function and $c \in [0,1]$ is a consistency bonus rewarding trajectories that behave as the hypothesis predicts. The weights $(\alpha_1, \alpha_2, \alpha_3) = (0.3, 0.3, 0.4)$ balance distance, improvement, and convergence contributions; normalization constants $(\tau_d, \tau_{\Delta}, \tau_{\kappa}) = (15, 8, 5)$ scale each statistic to comparable ranges; consistency coefficients $(\beta_0, \beta_1) = (0.6, 0.4)$ modulate the hypothesis bonus; and integration steps $T = 50$, all determined via cross-validation. This multi-statistic scheme enables compositional discrimination where manifold distance provides baseline signal while trajectory dynamics reveal fine-grained binding compatibility.
 
\textbf{Discriminative mechanism.}
The key insight is that binders and decoys respond differently to hypothesis-conditioned flow even when starting from their actual embeddings. For binders, the hypothesis generator outputs \texttt{CONVERGE} directives and the velocity field produces smooth trajectories maintaining proximity to the binder manifold, yielding high consistency bonus and positive improvement. For decoys, the hypothesis generator outputs \texttt{REJECT} directives and the velocity field produces trajectories that fail to converge--either because the embedding is geometrically incompatible with the binder subspace, or because \texttt{REJECT} conditioning explicitly discourages convergence. This creates a self-consistent discriminative loop where the hypothesis encodes prediction and the trajectory validates it.
 

\textbf{Flow trigger conditions.}
To balance computational cost against discriminative benefit, we cache hypothesis generation for $N_{\text{train}}$ representative pairs (equal binders and decoys) during training, storing parsed directives and LLM hidden states for downstream use. Pairs are selected via stratified sampling to ensure coverage across embedding geometry regions (binned by cosine similarity and L2 distance) preventing bias toward high-density manifold areas. The coverage threshold (cross-validated) balances representation space coverage against computational cost. At inference, pairs not in cache receive rule-based hypotheses derived from embedding features, maintaining discriminative signal without per-pair LLM queries. Flow-scored predictions are marked \texttt{HESFM\_SCORED} in the audit trail with full provenance for post-hoc analysis of directive effectiveness across embedding geometries.

\begin{algorithm}[h]
\caption{Hypothesis-Guided Entropy Regularized Tree of Thoughts Search}
\label{alg:tot}
\begin{algorithmic}[1]
\REQUIRE Receptor $P_R$, candidates $\mathcal{L}$, depth limit $D$, node budget $M$, branch factor $B$, temperature $\tau_0$, pruning threshold $\varepsilon_1, \varepsilon_2$
\STATE Initialize visited $\mathcal{V} \gets \emptyset$, best $(\ell^*, S^*) \gets (\varnothing, -\infty)$, directive history $\mathcal{D} \gets \emptyset$; push root to stack

\COMMENT{\textit{Phase 1: Hypothesis-guided pre-filtering}}
\FOR{each $\ell \in \mathcal{L}$}
  \STATE Compute embedding features $\mathbf{f}_\ell = (\text{cosine\_sim}, \text{prod\_mean}, \ldots)$ from $(\mathbf{e}_R, \mathbf{e}_\ell)$
  \STATE Generate directive $d_\ell \in \{\texttt{PRIORITIZE}, \texttt{EXPLORE}, \texttt{DEPRIORITIZE}, \texttt{SKIP}\}$ via finetuned LLM
  \STATE Assign priority $w_\ell \gets w_{d_\ell} \cdot c_\ell$ where $c_\ell$ is confidence
\ENDFOR
\STATE $\mathcal{L}' \gets \text{top-}k$ candidates by priority $w_\ell$

\COMMENT{\textit{Phase 2: Entropy-regularized tree search}}
\WHILE{stack nonempty \AND $|\mathcal{V}| < M$}
  \STATE Pop state $(P_R, \mathcal{V}, V_\mathrm{prev}, \Delta V_\mathrm{prev}, d_\mathrm{parent})$
  \STATE Compute adaptive temperature $\tau_t$ from directive history $\mathcal{D}$
  \STATE Compute $Q(x, a) = \lambda_V V(\phi(P_R, a)) + \lambda_h w_a$ for $a \in \mathcal{L}' \setminus \mathcal{V}$
  \STATE Let $\mathcal{A}_B$ be top $B$ candidates by $Q$; sample sequence from $\pi^*(a|x) \propto \exp(Q/\tau_t)$
  \FOR{each action $a$ in sampled sequence}
    \STATE Compute score vector $\mathbf{s}_a$, value $V_t$, differences $\Delta V_t$, $\Delta^2 V_t$
    \STATE Update $\mathcal{D} \gets \mathcal{D} \cup \{d_a\}$
    \IF{$\Delta V_t < \varepsilon_1$ \AND $\Delta^2 V_t \leq \varepsilon_2$ \AND $d_a \in \{\texttt{DEPRIORITIZE}, \texttt{SKIP}\}$}
      \STATE \textbf{continue} \COMMENT{Hypothesis-aware pruning}
    \ENDIF
    \IF{$d_a = \texttt{EXPLORE\_TENSION}$ \OR $\text{score\_range}(\mathbf{s}_a) > \gamma$}
      \STATE Generate hypothesis $h_a$ with directive $\texttt{CONVERGE}$ or $\texttt{REJECT}$
      \STATE Run embedding-space flow: $\mathbf{e}_\ell^{(T)} = \int_0^1 v_\theta(\mathbf{e}_\ell^{(t)}, t; \mathbf{e}_R, h_a) \, dt$
      \STATE Compute flow score from trajectory: final distance, improvement, convergence
      \STATE $V_t \gets \text{combine}(V_t, \text{flow\_score}, \text{consistency}(h_a))$
    \ENDIF
    \STATE Update best if $V_t > S^*$; record $(a, \mathbf{s}_a, d_a, V_t)$ in tree $\mathcal{T}$; add $a$ to $\mathcal{V}$; push child
  \ENDFOR
\ENDWHILE
\STATE Generate explanations for top $k$ nodes via finetuned LLM
\RETURN Best ligand(s) $\ell^*$, tree $\mathcal{T}$ with scores, directives, and explanations
\end{algorithmic}
\end{algorithm}


\section{Experiments}\label{sec:experiments}

We evaluate Protein Thoughts on its core objective: hypothesis-guided PPI discovery with interpretable evidence. We first validate the value function components (Sections~\ref{sec:exp_decomp}--\ref{sec:exp_screen}), then assess the full discovery pipeline on large-scale search (Section~\ref{sec:exp_discovery}) and isolate the contribution of hypothesis-conditioned flow matching (Section~\ref{sec:exp_flow}).

\subsection{Score Decomposition and Value Function Validation}
\label{sec:exp_decomp}

We validate that the four-score decomposition captures distinct interaction regimes through controlled benchmarks (Appendix~\ref{app:examples}). The barnase--barstar system (PDB: 1BRS, $K_d \approx 10^{-14}$\,M) represents one of the highest affinity protein--protein interactions known \citep{buckle1994}. Barnase is an extracellular ribonuclease from \emph{Bacillus amyloliquefaciens} and barstar is its intracellular inhibitor essential for protecting the host from barnase toxicity during biosynthesis.

We used chain A of 1BRS (barnase, 110 residues) as the receptor and constructed a candidate pool of 50 protein chains including the true binding partner (1BRS\_B, barstar) and 49 negative controls drawn from unrelated PDB structures. Negative controls were selected to span diverse folds and sizes. Protein Thoughts correctly ranked 1BRS\_B as the top candidate with value $V = 0.78$.

The decomposed scores revealed the biological basis for this prediction. Sequence similarity of 0.98 reflects coevolutionary history where barnase and barstar have maintained compatible interfaces over millions of years. Structural complementarity of 0.99 indicates near perfect geometric fit where barstar's binding loop inserts into barnase's active site cleft. Interface balance of 0.99 shows both partners contribute comparably to the 1590\,\AA$^2$ buried interface. Chemical compatibility of 0.15 is modest, reflecting that binding is dominated by electrostatic steering and shape complementarity rather than optimized residue residue contacts at every position. The exceptionally fast association rate $k_\mathrm{on} \approx 10^9$\,M$^{-1}$s$^{-1}$ arises from charged interface residues that accelerate association rather than maximize equilibrium affinity.

The generated explanation correctly identified this as a geometry driven interaction:

\medskip
\noindent\fbox{\parbox{0.96\columnwidth}{\small
\textbf{Candidate: 1BRS\_B \quad Value: 0.78}\\[2pt]
\texttt{<thinking>} \textbf{Alignment (0.98):} High sequence similarity indicates coevolutionary binding compatibility. \textbf{Structural (0.99):} Near perfect geometric complementarity supports lock and key fit. \textbf{Contact ratio (0.99):} Both partners bury comparable surface area, signature of stable heterodimers. \textbf{Chemistry (0.15):} Modest, but barnase--barstar binding is driven by electrostatic steering and shape complementarity. The charged interface accelerates association rather than maximizing equilibrium affinity. \textbf{Synthesis:} Three signals agree strongly. Low chemistry is a mechanistic signature, not a contradiction.
\texttt{</thinking>}\\[2pt]
\texttt{<conclusion>} \textbf{Promote with high confidence.} Shape complementarity and interface balance drive binding. Modest chemistry reflects electrostatic steering.
\texttt{</conclusion>}\\[2pt]
\texttt{<next\_steps>} Inspect Asp35--Lys27 electrostatic pair. Alanine scan predicted hotspots. Run docking refinement.
\texttt{</next\_steps>}
}}
\medskip

We additionally evaluate an antibody--antigen system (PDB: 4M5Z), which exhibits characteristic tension: near-zero structural similarity ($s_\text{struct}=0.001$) together with high interface balance ($s_\text{contact}=0.94$), establishing $T = s_\text{contact} - s_\text{struct} > 0$ as a signature for CDR-driven recognition. All 25 negative controls receive $V < 0.25$ (0\% FPR in the top-5; see Appendix~\ref{app:negative_controls}). Table~\ref{tab:score_decomposition} summarizes the decomposed evidence.

\begin{table}[h]
\caption{\textbf{Decomposed evidence for benchmark complexes.} Barnase--barstar shows concordant structural scores; antibody--antigen exhibits tension (low $s_\text{struct}$, high $s_\text{contact}$).}
\label{tab:score_decomposition}
\centering\small
\begin{tabular}{lcccccl}
\toprule
\textbf{Pair} & $s_\mathrm{seq}$ & $s_\mathrm{struct}$ & $s_\mathrm{contact}$ & $s_\mathrm{chem}$ & $V$ & \textbf{Verdict} \\
\midrule
1BRS\_A -- 1BRS\_B & 0.98 & 0.99 & 0.99 & 0.15 & 0.78 & Promote \\
1BRS\_A -- 5UH9\_B & 0.23 & 0.00 & 0.45 & 0.15 & 0.21 & Deprioritize \\
4M5Z\_A -- 4M5Z\_H & 0.38 & 0.00 & 0.94 & 0.17 & 0.37 & Keep (tension) \\
\bottomrule
\end{tabular}
\end{table}

\textbf{Supervised PPI prediction.} On the SHS27k and SHS148k benchmarks, which are subsets derived from STRING~\citep{szklarczyk2021string} and evaluated using the DFS 70/10/20 split protocol introduced by GNN-PPI~\citep{lv2021learning} and subsequently adopted in later benchmark studies, PPIProjectedNet achieves $85.60 \pm 0.80$ and $91.08 \pm 0.19$ Micro-F1 respectively, outperforming prior methods (Table~\ref{tab:main_ppi_results}). Results on the full STRING homo sapiens benchmark are provided in Appendix~\ref{app:ppi_benchmarks}. The interpretable bottleneck incurs no accuracy penalty.

We compare against both non-LLM and LLM-based baselines. Non-LLM methods include GNN-PPI~\cite{lv2021learning}, which encodes proteins as residue graphs with interaction-type-aware aggregation, and PIPR~\cite{chen2019multifaceted}, which uses a residue-level co-attentive siamese network. LLM-based baselines include ProLLM~\cite{jin2023prollm}, which fine-tunes a large protein language model with interaction-type-conditioned instruction tuning, and LLAPA~\cite{zhou2024llapa}, which extends this with a lightweight adapter architecture on top of frozen protein language model representations.

Training PPIProjectedNet uses a multi-task loss combining binary classification and bottleneck alignment:
\begin{equation}
    \mathcal{L} =
    \underbrace{\mathcal{L}_\text{BCE}(\hat{y},\, \tilde{y})}_{\text{interaction loss}}
    + \;\lambda\,\underbrace{\|\mathbf{z} - \mathbf{s}\|_2^2}_{\text{bottleneck alignment}}
\end{equation}
where $\tilde{y} = y(1-\epsilon) + 0.5\epsilon$ with $\epsilon=0.10$
and $\lambda=0.5$. Here $\mathbf{z}$ denotes the four-dimensional interpretable bottleneck derived in the PPIProjectedNet architecture, and $\mathbf{s}$ are the corresponding biological target scores.

Training uses AdamW with learning rate $3\times 10^{-4}$,
weight decay $5\times 10^{-3}$, cosine annealing with warmup,
and early stopping based on validation loss.

\begin{table}[h]
\centering
\caption{Micro-F1 scores (\%) on SHS27k and SHS148k under the DFS 70/10/20 split protocol introduced by~\cite{lv2021learning}. \textbf{Bold} indicates the best result per column. Methods are grouped by type: \emph{Non-LLM} methods use sequence or graph features without a pretrained language model; \emph{LLM-based} methods build on protein language model representations.}
\label{tab:main_ppi_results}
\small
\begin{tabular}{llcc}
\toprule
\textbf{Type} & \textbf{Method} & \textbf{SHS27k} & \textbf{SHS148k} \\
\midrule
\multirow{2}{*}{Non-LLM} & GNN-PPI & 67.43 & 64.97 \\
& PIPR & 52.19 & 61.38 \\
\midrule
\multirow{2}{*}{LLM-based} & ProLLM & 85.32 & 87.66 \\
& LLAPA & 69.54 & 73.93 \\
\midrule
Ours & \textbf{PPIProjectedNet} & $\mathbf{85.60}$ & $\mathbf{91.08}$ \\
\bottomrule
\end{tabular}
\end{table}

\textbf{Mutation sensitivity.} On SKEMPI v2~\citep{jankauskaite2019}, the model consistently separates stabilizing from destabilizing mutations (Figure~\ref{fig:skempi_results}). For each SKEMPI entry we reconstruct wild type sequences and apply the mutation to generate the mutated pair. From wild type $(A, B)$ and mutated $(A', B')$ we compute score differences $\Delta s = s(A', B') - s(A, B)$. Additional mutation descriptors include amino acid substitution indicators, physicochemical property differences, and local sequence context.

The model trains as a multi task predictor with three outputs: a binary classifier predicting whether the mutation improves binding, a regression head estimating $\Delta\Delta G$, and an auxiliary flow head predicting $\Delta s$. The classifier is the primary objective while regression and flow heads act as auxiliary tasks encouraging structured perturbation learning. Grouped train/validation/test splits by protein complex identifier prevent information leakage.

The ROC curve in Figure~\ref{fig:skempi_results} shows consistent separation between stabilizing and destabilizing mutations, indicating that learned features capture relevant structural and chemical signals. The flow matching head achieves low MSE across all four score channels, confirming that mutations can be characterized as structured perturbations in the interpretable score space.

\subsection{Virtual Screening}
\label{sec:exp_screen}

To complement classification metrics, we evaluate PPIProjectedNet in a
virtual screening setting on the SHS27k test split.
The receptor in this experiment is \textit{SPOP} (Speckle-type POZ
protein, \texttt{9606.ENSP00000240327}), a BTB/POZ domain protein that
acts as a substrate-recognition adaptor for the Cullin-3 ubiquitin
ligase complex and modulates transcriptional repression via DAXX.
Crucially, \textit{SPOP} is the most commonly point-mutated gene in
human primary prostate cancer, with mutations implicated in
approximately 15\% of all prostate cancer cases, making its interaction
network of direct clinical relevance.

We construct a 500-candidate pool containing all five confirmed binding
partners recorded in SHS27k together with 495 randomly sampled
non-interacting decoys, then rank all 500 candidates by predicted
binding probability without applying any threshold.

\begin{figure}[h]
\centering
\includegraphics[width=\linewidth]{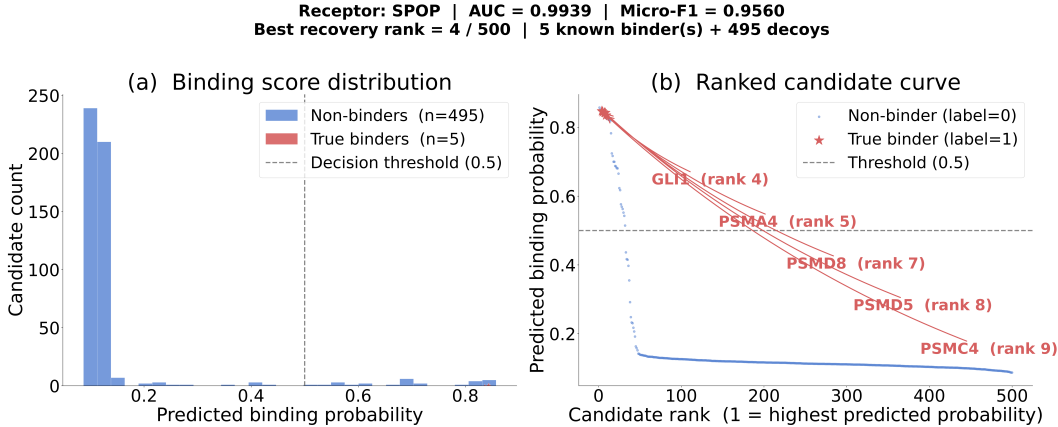}
\caption{Virtual screening for SPOP (500 candidates). \textbf{(a)} Probability distributions for non-binders (blue) vs.\ confirmed partners (red). \textbf{(b)} All five binders (red stars) recovered within rank 9 (AUC=0.994).}
\label{fig:virtualscreen}
\end{figure}

All five confirmed interaction partners are recovered within the top~9
ranks out of 500: \textit{GLI1} (rank~4), a transcription
factor and downstream effector of the Hedgehog signalling pathway;
and four subunits of the 26S proteasome complex, \textit{PSMA4}
(rank~5), \textit{PSMD8} (rank~7), \textit{PSMD5} (rank~8), and
\textit{PSMC4} (rank~9), consistent with SPOP's established role in
directing substrates to the proteasomal degradation machinery.
Recovery of all five known partners within the top~9 candidates
corresponds to a mean enrichment factor of ${\approx}56\times$ above
random expectation, with the model achieving AUC\,=\,0.994 and
Micro-F1\,=\,0.956 (Figure~\ref{fig:virtualscreen}).
These results confirm that PPIProjectedNet recovers biologically
meaningful interaction partners near the top of a large candidate pool.

\subsection{Hypothesis-Guided Discovery on SHS148k}
\label{sec:exp_discovery}

We evaluate the full discovery pipeline on SHS148k, where each receptor faces ${\sim}5{\times}10^3$ candidates. We compare the entropic ToT baseline against hypothesis-guided ToT with Qwen-generated directives, adaptive temperature scheduling, pruning, and selective flow triggering.

Across 10 receptors, hypothesis-guided ToT reduces mean best-binder rank from 47.70 to 11.20 ($\Delta$rank $= +36.50$), improving in 8/10 trials and achieving rank 1 in 7/10 (Table~\ref{tab:tot_pertrial}). The single failure case (trial 2) involves a receptor with one known binder where ambiguous embedding features misled the hypothesis generator, highlighting that hypothesis quality remains a bottleneck for difficult edge cases.

\begin{table}[h]
\caption{\textbf{Discovery results on SHS148k} (10 receptors, ${\sim}5{\times}10^3$ candidates each). $\Delta$rank $=$ baseline $-$ Qwen (positive = improvement).}
\label{tab:tot_pertrial}
\centering\small
\begin{tabular}{clcccc}
\toprule
\textbf{Trial} & \textbf{Receptor} & \textbf{Baseline} & \textbf{Qwen} & \textbf{$\Delta$rank} & \textbf{\#Bind} \\
\midrule
1  & \texttt{ENSP00000297837} & 55 & \textbf{1} & $+54$ & 11 \\
2  & \texttt{ENSP00000254325} & 40 & 55 & $-15$ & 1 \\
3  & \texttt{ENSP00000278765} & 55 & \textbf{1} & $+54$ & 1 \\
4  & \texttt{ENSP00000217086} & 51 & 49 & $+2$ & 1 \\
5  & \texttt{ENSP00000240361} & 55 & \textbf{1} & $+54$ & 1 \\
6  & \texttt{ENSP00000228437} & 1  & \textbf{1} & $0$ & 3 \\
7  & \texttt{ENSP00000263464} & 55 & \textbf{1} & $+54$ & 1 \\
8  & \texttt{ENSP00000304930} & 55 & \textbf{1} & $+54$ & 1 \\
9  & \texttt{ENSP00000303015} & 55 & \textbf{1} & $+54$ & 5 \\
10 & \texttt{ENSP00000301305} & 55 & \textbf{1} & $+54$ & 2 \\
\midrule
\textbf{Mean} & & 47.70 & \textbf{11.20} & $\mathbf{+36.50}$ & -- \\
\bottomrule
\end{tabular}
\end{table}

\subsection{Effect of Hypothesis-Conditioned Flow Matching}
\label{sec:exp_flow}

We isolate H-ESFM's contribution by comparing rankings before and after flow on 1351-candidate sets from SHS27k (Table~\ref{tab:flow_ablation}). Raw L2 distance $d_0$ in ESM-2 space is near-random (rank 339.30, AUC 0.50). The value function $V_\phi$ alone substantially recovers signal (rank 38.30, AUC 0.93). H-ESFM further improves to rank 26.70 with AP nearly doubling (0.064 $\to$ 0.123), confirming that hypothesis-conditioned flow reshapes embedding geometry beyond what $V_\phi$ achieves. Flow improves best-binder rank in 8/10 receptors (Wilcoxon $p = 0.027$).

\begin{table}[h]
\caption{\textbf{Ablation of H-ESFM} (10 receptors, 1351 candidates each). H-ESFM improves ranking beyond $V_\phi$ alone, with AP nearly doubling.}
\label{tab:flow_ablation}
\centering\small
\begin{tabular}{lcccc}
\toprule
\textbf{Condition} & \textbf{Best Rank}$\downarrow$ & \textbf{AUC} & \textbf{AP} & \textbf{Mean Rank}$\downarrow$ \\
\midrule
Initial L2 $d_0$ & 339.30 & 0.499 & 0.041 & 677.94 \\
$V_\phi$ only & 38.30 & 0.928 & 0.064 & 99.88 \\
H-ESFM $d_T$ & \textbf{26.70} & \textbf{0.931} & \textbf{0.123} & \textbf{95.62} \\
\bottomrule
\end{tabular}
\end{table}

\begin{figure}[h]
\centering
\includegraphics[width=0.5\linewidth]{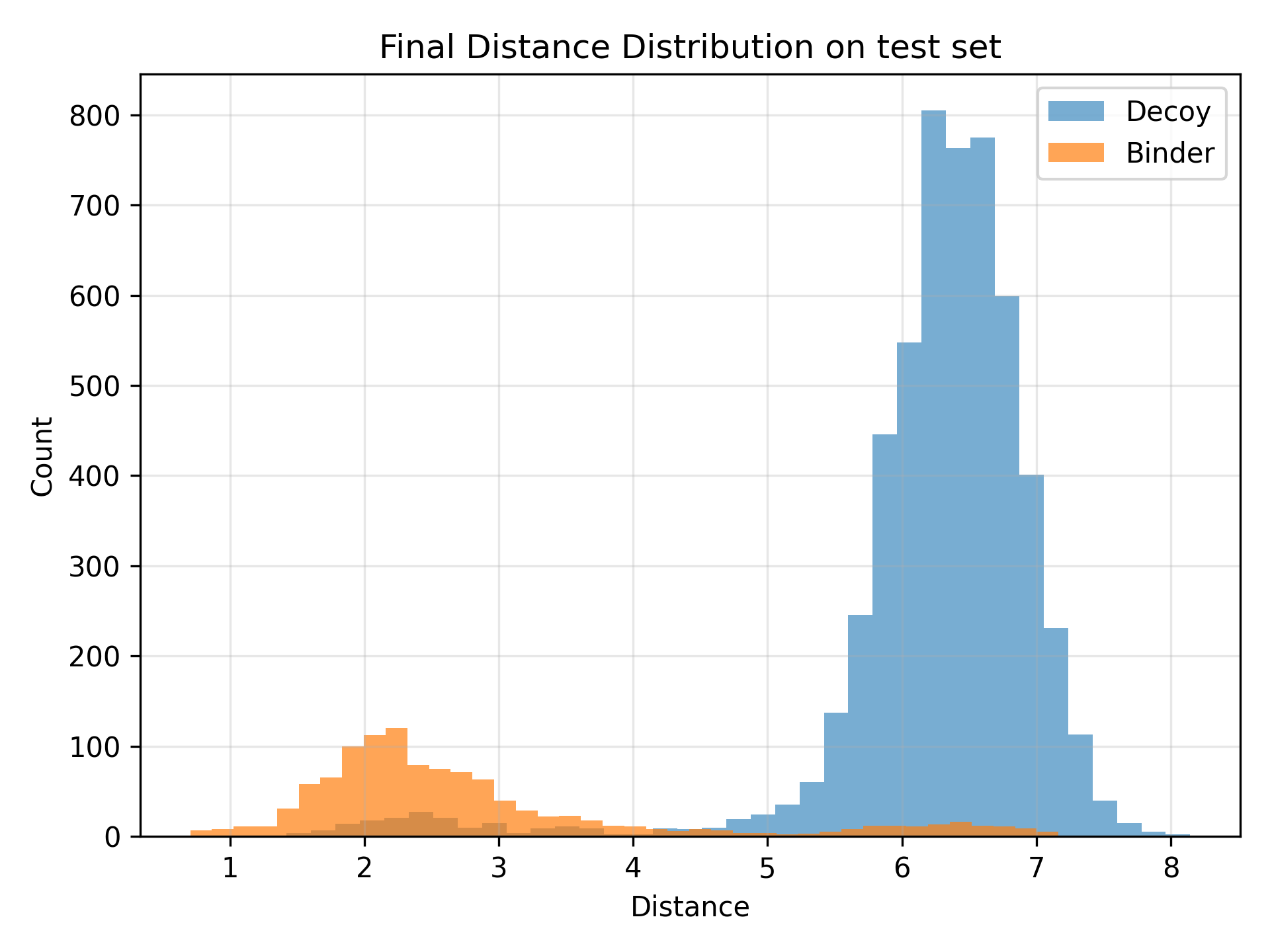}
\caption{\textbf{Final distance after flow.} Flow pushes decoys from the binder manifold; imperfect separation suggests room for improvement with more training data.}
\label{fig:flow_distance_test}
\end{figure}

Figure~\ref{fig:flow_distance_test} shows that flow pushes many decoys away from the binder manifold while improving a subset of candidates; incomplete separation indicates that larger-scale training data would further improve generalization.

Table~\ref{tab:flow_analysis} breaks down H-ESFM by perturbation strategy. Contact maximization and interface approach achieve the highest success rates (89\%, 84\%), consistent with thermodynamic drive to bury surface area. Backbone flexibility triggers less frequently but yields the largest improvement when successful (+0.11), resolving harder cases where induced-fit adjustments unlock binding potential.

\subsection{Full Framework Evaluation}
\label{sec:full_eval}
We evaluate the complete framework on 10 held-out receptors, comparing tree search alone, tree search with LLM-guided hypothesis generation (ToT+Qwen), and the full system with H-ESFM integrated at each expansion step (Table~\ref{tab:full_eval}). The full framework achieves a mean best-rank of 2.20, a $4.2\times$ improvement over tree search alone (9.30) and ahead of ToT+Qwen (9.00), with H-ESFM improving 8/10 receptors (Wilcoxon $p = 0.020$). Confirming that geometric refinement via flow matching is important for ranking improvement; full per-receptor results appear in Appendix~\ref{sec:app_full_eval}.

\begin{table}[h]
\caption{\textbf{Full framework evaluation} (10 receptors). Adding H-ESFM flow to tree search drives the largest ranking gains; LLM hypothesis generation alone contributes marginally.}
\label{tab:full_eval}
\centering\small
\begin{tabular}{lcc}
\toprule
\textbf{System} & \textbf{Mean Best Rank}$\downarrow$ & \textbf{Receptors Improved} \\
\midrule
Tree search (baseline) & 9.30 & -- \\
ToT + Qwen hypothesis & 9.00 & 2/10 \\
ToT + Qwen + H-ESFM & \textbf{2.20} & \textbf{8/10} \\
\bottomrule
\end{tabular}
\end{table}

\textbf{Remark.} When PPIProjectedNet (baseline tree search) already achieves a high baseline rank, the margin for further improvement narrows for any method: discriminating among candidates already close to the binder manifold is strictly harder, and gains from both Hypothesis-Guided Discovery and H-ESFM are more marginal in such cases.

\textbf{Summary.} The value function provides strong baseline ranking with interpretable scores. Hypothesis-guided ToT achieves 76\% rank improvement over entropic baseline by leveraging LLM-generated directives for adaptive search. H-ESFM further refines rankings by reshaping embedding geometry under hypothesis conditioning, with AP doubling relative to $V_\phi$ alone. Together, these components enable efficient, interpretable PPI discovery at scale.

\section{Discussion}
\label{sec:discussion}
\label{app:discussion}

\paragraph{Interpretability as a design principle.}
Protein Thoughts demonstrates that interpretability can be built into computational biology pipelines from the ground up rather than added as an afterthought~\citep{chen2024interpretable}. The four decomposed signals were chosen not for maximal predictive power but for interpretability where each corresponds to a biological concept that practitioners routinely consider, and each can be questioned using domain expertise~\citep{novakovsky2023explainable}. When the system is correct, the audit trail reveals the reason. On the other hand, when wrong, it reveals how both outcomes support scientific learning in ways that black-box predictions cannot provide~\citep{vig2021bertology}.
A learned end-to-end model might achieve marginally higher accuracy by discovering subtle correlations invisible to hand-crafted features. However, such a model would sacrifice the transparency that enables scientific engagement~\citep{simon2024language}. When Protein Thoughts predicts that two proteins bind, the decomposition reveals which signals agreed and which tensions were informative. When the prediction fails, the decomposition reveals which evidence was misleading. This transparency distinguishes our approach from black-box rankers and aligns with recent calls for interpretable methods that yield biological insight rather than mere prediction~\citep{chen2024interpretable,novakovsky2023explainable}.

\paragraph{Tension as mechanistic signal.}
Many of our examples including the antibody–antigen case illustrates a deeper insight which is that the signal disagreement is not noise to be minimized but mechanistic structure to be preserved~\citep{selaculang2013structural}. The tension profile such as high $s_\mathrm{contact}$ paired with near-zero $s_\mathrm{struct}$, correctly characterizes CDR-driven binding~\citep{rives2023igfold}. A system that averages away this tension loses precisely the information needed for mechanistic understanding. The tension score $T = s_\mathrm{contact} - s_\mathrm{struct}$ thus serves as a diagnostic for interaction type, enabling the system to apply different reasoning templates for enzyme–inhibitor versus antibody–antigen recognition~\citep{zhao2024mgppi}.

\paragraph{Hypothesis-conditioned flow matching.}
Our approach operates in embedding space guided by LLM-generated hypotheses grounded in biochemical intuition, e.g., contact maximization, interface approach, rigid body optimization, rotational sampling, and backbone flexibility—each encoding domain knowledge about protein binding mechanisms. Contact maximization reflects thermodynamic drive to bury surface area; backbone flexibility captures local induced fit. Together, these strategies span conformational changes relevant to most PPIs without requiring expensive molecular dynamics, enabling exploration of induced-fit mechanisms that score-space methods cannot capture.

\paragraph{Implications for drug discovery.}
In therapeutic development, identifying and validating protein targets remains a costly bottleneck with high attrition rates often attributed to incomplete mechanistic understanding \citep{hughes2011,paul2010}. By decomposing binding evidence into biologically meaningful signals, Protein Thoughts enables medicinal chemists to prioritize candidates not merely by predicted affinity but by the nature of the interaction. A candidate showing high interface balance but modest chemical compatibility may warrant optimization of specific contact residues. One exhibiting tension between structural fit and evolutionary signal may suggest allosteric or transient binding requiring distinct assay strategies \citep{arkin2014,wells2007}. 
For de novo protein design \citep{huang2016,watson2023}, the decomposed scoring provides direct feedback. A designed variant that improves structural complementarity but disrupts interface balance signals a geometric mismatch requiring backbone adjustment. When integrated with generative protein design methods \citep{hsu2022,liu2025mctd}, Protein Thoughts can serve as an interpretable discriminator providing actionable mechanistic feedback for subsequent design rounds. 

\paragraph{Limitations.}
Scoring components use simple biophysical proxies. Integration with learned binding free energy predictors \citep{weng2019,wang2019} could improve accuracy at the cost of computation. Structural scoring assumes fixed conformations, potentially failing for intrinsically disordered proteins \citep{wright2015,uversky2019}. Flow matching provides a partial remedy but operates through discrete perturbation strategies rather than continuous dynamics. Systematic evaluation on larger benchmarks with rigorous cross validation remains future work. The approximately 20\% explanation hallucination rate means explanations should be treated as hypotheses for critique, not authoritative statements.

\section{Conclusion}
We presented Protein Thoughts, a framework for interpretable PPI discovery that decomposes binding evidence into four biologically meaningful signals, structures exploration via hypothesis-guided Tree-of-Thoughts search, and resolves score tension through embedding-space flow matching. PPIProjectedNet achieves 85.60, 91.08, and 98.05 Micro-F1 on SHS27k, SHS148k, and STRING, while hypothesis-guided discovery reduces mean best-binder rank from 47.7 to 11.2 by screening 500 candidates in under 30 seconds (2000$\times$ faster than AlphaFold-Multimer). Controlled benchmarks validate interpretability: barnase–barstar is promoted on concordant geometry, while antibody–antigen recognition is captured through characteristic score tension. Limitations include simplified biophysical proxies and hypothesis-generator brittleness on ambiguous embeddings. As computational biology advances, systems that expose reasoning will be essential. Protein Thoughts offers one path forward.

\paragraph{Acknowledgments.} This material is based upon work supported by the Google Cloud Research Credits program, award GCP19980904. We are also grateful to Marina Garassino, Tobin Sosnick, Yitan Zhu for their encouragement and valuable insights during the early stages of this project.

{\small
\bibliographystyle{plainnat}

}

\appendix

\section{Protein Representation and Scoring Functions}
\label{sec:scoring}

\paragraph{Protein representation.}
Proteins are represented as tuples $P = (\mathrm{id}, s, \boldsymbol{\phi}, c)$ where $\mathrm{id}$ is a unique identifier, $s \in \Sigma^*$ is the amino acid sequence over the 20-letter alphabet $\Sigma = \{\mathrm{A,C,D,E,F,G,H,I,K,L,M,N,P,Q,R,S,T,V,W,Y}\}$, $\boldsymbol{\phi}(P) \in \mathbb{R}^{64 \times 6}$ is a fixed-length physicochemical profile computed from $s$ (defined below), and $c \in \mathbb{R}_{>0}$ estimates the number of interface-capable residues. Sequence embeddings are obtained from ESM-2~\citep{lin2023evolutionary} (35M parameter variant), producing per-protein representations $\mathbf{e} \in \mathbb{R}^{480}$ used as input to PPIProjectedNet alongside the four scalar scores defined below.

\paragraph{Sequence similarity.}
$k$-mer Jaccard similarity with $k=3$ is computed over overlapping trimers of each cleaned sequence:
\begin{equation}
s_{\mathrm{seq}} = \frac{|A_3 \cap B_3|}{|A_3 \cup B_3|} \in [0,1],
\end{equation}
where $A_3$ and $B_3$ are the 3-mer sets of $P_R$ and $P_L$ respectively. If either sequence has length $<3$, $s_{\mathrm{seq}} = 0$. High values reflect shared sequence motifs and coevolutionary history~\citep{marks2011,dejuan2013,cong2019}.

\paragraph{Structural complementarity.}
Each sequence of length $n$ is mapped to a per-residue property matrix $\mathbf{F} \in \mathbb{R}^{n \times 6}$, where the six columns encode hydrophobicity, polarity, formal charge, aromaticity, size, and molecular weight (normalized by 200~Da) using a fixed amino acid property table. This variable-length matrix is resampled to $B = 64$ uniformly spaced positions via piecewise linear interpolation independently per channel. Letting $x_i = i/(n-1)$ denote original sample positions and $\hat{x}_b = b/(B-1)$ the target positions, the resampled value at bin $b$ for channel $j$ is:
\begin{equation}
\boldsymbol{\phi}_j(P)[b] = \mathbf{F}_{i,j} + \bigl(\hat{x}_b - x_i\bigr)\frac{\mathbf{F}_{i+1,j} - \mathbf{F}_{i,j}}{x_{i+1} - x_i}, \quad i = \max\bigl\{i' : x_{i'} \le \hat{x}_b\bigr\},
\end{equation}
yielding $\boldsymbol{\phi}(P) \in \mathbb{R}^{64 \times 6}$. The structural score is the shifted cosine similarity of the flattened profiles $\mathbf{A} = \mathrm{vec}(\boldsymbol{\phi}(P_R)),\, \mathbf{B} = \mathrm{vec}(\boldsymbol{\phi}(P_L)) \in \mathbb{R}^{384}$:
\begin{equation}
s_{\mathrm{struct}} = \frac{1}{2}\!\left(\frac{\mathbf{A} \cdot \mathbf{B}}{\|\mathbf{A}\|\,\|\mathbf{B}\| + \epsilon} + 1\right) \in [0,1],
\end{equation}
with $\epsilon = 10^{-12}$. The $+1$ shift maps cosine similarity from $[-1,1]$ to $[0,2]$ before halving, ensuring non-negativity. This proxy captures global physicochemical shape complementarity without requiring 3D coordinates~\citep{mian1991}.

\paragraph{Interface balance.}
A residue is considered interface-capable if it belongs to the set $\mathcal{I} = \{\mathrm{D,E,K,R,H,N,Q,S,T,Y,W,F}\}$, which is enriched at protein--protein interfaces~\citep{janin1990,bogan1998,chakrabarti2002}. The estimated interface residue count is:
\begin{equation}
c(P) = |P|\,(0.20 + 0.45\,f_{\mathcal{I}}),
\end{equation}
where $f_{\mathcal{I}}$ is the fraction of residues in $\mathcal{I}$. The interface balance score is:
\begin{equation}
s_{\mathrm{contact}} = \frac{\min(c_R,\, c_L)}{\max(c_R,\, c_L)} \in (0,1],
\end{equation}
equal to 1 when both partners contribute symmetrically to the interface, consistent with the observation that stable heterodimeric complexes typically bury comparable surface area on each partner~\citep{locontechothia1999}.

\paragraph{Chemical compatibility.}
A symmetric $20\times 20$ interaction matrix $M \in \mathbb{R}^{20\times 20}$ is pre-computed over the standard amino acid alphabet using physicochemical rules~\citep{janin1990,chakrabarti2002}. For each ordered pair $(a,b)$, the raw entry is:
\begin{align}
\tilde{M}_{ab} &= 0.6\cdot\mathbbm{1}[\mathrm{hyd}_a \wedge \mathrm{hyd}_b]
               + 0.4\cdot\mathbbm{1}[\mathrm{aro}_a \wedge \mathrm{aro}_b]
               + 0.9\cdot\mathbbm{1}[q_a q_b < 0] \notag \\
              &\quad - 0.5\cdot\mathbbm{1}[q_a q_b > 0]
               + 0.3\cdot\mathbbm{1}[\mathrm{pol}_a \wedge \mathrm{pol}_b \wedge q_a{=}0 \wedge q_b{=}0] \notag \\
              &\quad - 0.01\,|\mathrm{MW}_a - \mathrm{MW}_b|,
\end{align}
where $q_a$ is the formal charge and $\mathrm{MW}_a$ the molecular weight in Da. The matrix is symmetrized as $M \leftarrow (\tilde{M} + \tilde{M}^\top)/2$. Sequences are truncated to 160 residues before scoring. The chemical compatibility score is:
\begin{equation}
s_{\mathrm{chem}} = \mathrm{clip}\!\left(\frac{\bar{M}(P_R, P_L) + 1.5}{3.0},\ 0,\ 1\right) \in [0,1],
\end{equation}
where $\bar{M}(P_R, P_L) = \frac{1}{|P_R||P_L|}\sum_{i,j} M_{P_R[i],\, P_L[j]}$. The affine map places the typical raw range $[-1.5, 1.5]$ onto $[0,1]$, with clipping handling outliers. If either sequence is empty after cleaning, $s_{\mathrm{chem}}$ defaults to $0.5$.

\section{Further Experimental Validation and Details}

We evaluated Protein Thoughts on a series of PPI tasks spanning diverse interaction regimes including enzyme inhibitor complexes with high affinity shape driven binding, antibody antigen recognition with localized epitope paratope contacts, negative controls with no known interaction, and large scale evaluation on the seq\_ppi dataset. Our goals were to assess whether the system recovers known binding partners, whether the decomposed scores capture meaningful biological signals, whether learned weighting improves predictive accuracy, and whether the resulting explanations remain scientifically plausible.

\subsection{Example 1: Barnase--Barstar Recovery}
\label{app:examples}

The barnase--barstar system (PDB: 1BRS) represents one of the highest affinity protein--protein interactions known with $K_d \approx 10^{-14}$\,M \citep{buckle1994}. Barnase is an extracellular ribonuclease from Bacillus amyloliquefaciens and barstar is its intracellular inhibitor essential for protecting the host from barnase toxicity during biosynthesis.

\paragraph{Setup.}
We used chain A of 1BRS (barnase, 110 residues) as the receptor and constructed a candidate pool of 50 protein chains including the true binding partner (1BRS\_B, barstar) and 49 negative controls drawn from unrelated PDB structures. Negative controls were selected to span diverse folds and sizes.

\paragraph{Results.}
Protein Thoughts correctly ranked 1BRS\_B as the top candidate with value $V = 0.78$ (Table~\ref{tab:results}). The decomposed scores revealed the biological basis for this prediction. Sequence similarity of 0.98 reflects coevolutionary history where barnase and barstar have maintained compatible interfaces over millions of years. Structural complementarity of 0.99 indicates near perfect geometric fit where barstar's binding loop inserts into barnase's active site cleft. Interface balance of 0.99 shows both partners contribute comparably to the 1590\,\AA$^2$ buried interface. Chemical compatibility of 0.15 is modest, reflecting that binding is dominated by electrostatic steering and shape complementarity rather than optimized residue residue contacts at every position. The exceptionally fast association rate $k_\mathrm{on} \approx 10^9$\,M$^{-1}$s$^{-1}$ arises from charged interface residues that accelerate association rather than maximize equilibrium affinity.
\medskip

\subsection{Example 2: Antibody--Antigen Recognition}

We evaluated antibody 5J8 bound to pandemic influenza hemagglutinin (PDB: 4M5Z) \citep{hong2013}. Antibody antigen recognition presents a distinct challenge because antibodies achieve binding through localized CDR contacts without adopting the global fold of their targets. Standard structural similarity metrics that compare overall backbone geometry systematically fail for this interaction class.

\paragraph{Setup.}
Chain A (hemagglutinin HA1 subunit, 327 residues) served as the receptor. The candidate pool included the true antibody partner (4M5Z\_H, heavy chain) and 49 controls.

\paragraph{Results.}
The antibody chain 4M5Z\_H was recovered within the top 3 candidates with value $V = 0.37$. The decomposed scores revealed a characteristic tension profile (Table~\ref{tab:results}). Sequence similarity of 0.38 is moderate, reflecting partial conservation in CDR adjacent framework regions but no global homology between immunoglobulin and viral hemagglutinin folds. Structural complementarity of 0.001 is near zero, correctly reflecting that rigid body superposition of antibody and antigen produces poor geometric match. Interface balance of 0.94 is high, indicating both partners contribute substantially to the interface. Antibody heavy chains typically bury 600 to 900\,\AA$^2$ at the paratope and the HA1 epitope buries comparable area. Chemical compatibility of 0.17 is modest, reflecting that binding depends on specific CDR epitope contacts rather than globally favorable chemistry.

The tension between structural dissimilarity and interface balance is precisely what distinguishes antibody antigen recognition from enzyme inhibitor complexes:

\medskip
\noindent\fbox{\parbox{0.96\columnwidth}{\small
\textbf{Candidate: 4M5Z\_H \quad Value: 0.37}\\[2pt]
\texttt{<thinking>} \textbf{Alignment (0.38):} Moderate. CDR adjacent framework conservation but no global homology between immunoglobulin and HA folds. \textbf{Structural (0.001):} Near zero. Expected for antibody antigen recognition. Rigid body superposition is the wrong metric here. \textbf{Contact ratio (0.94):} High. Both partners contribute substantially. Antibody heavy chains bury 600 to 900\,\AA$^2$ at the paratope. \textbf{Chemistry (0.17):} Modest. Binding via specific CDR loop hotspots. \textbf{Tension:} Structural dissimilarity versus high interface balance. This pattern is the defining signature of antibody antigen recognition. Localized CDR contacts create stable binding despite global geometric divergence. The tension is not a contradiction. It is mechanistically informative.
\texttt{</thinking>}\\[2pt]
\texttt{<conclusion>} \textbf{Keep and explore.} Classic tension profile for antibody recognition. CDR loops engage a defined surface patch on HA1.
\texttt{</conclusion>}\\[2pt]
\texttt{<next\_steps>} Map CDR loops H1/H2/H3 onto HA1 receptor binding site. Run flexible docking. Compute buried surface per CDR loop.
\texttt{</next\_steps>}
}}
\medskip

The tension score $T = s_\mathrm{contact} - s_\mathrm{struct}$ is large and positive for all PDB antibody antigen complexes we examined, establishing it as a diagnostic for CDR driven binding. A method relying solely on structural similarity systematically misses this entire interaction class.

\begin{table}[t]
\caption{\textbf{Decomposed evidence for benchmark complexes.} Protein Thoughts recovers known binding partners while exposing the biological signals underlying each prediction. The barnase--barstar pair shows concordant high scores across structural signals. The antibody--antigen pair exhibits characteristic tension with near zero structural similarity but high interface balance.}
\label{tab:results}
\centering\small
\begin{tabular}{lcccccl}
\toprule
\textbf{Pair} & $s_\mathrm{seq}$ & $s_\mathrm{struct}$ & $s_\mathrm{contact}$ & $s_\mathrm{chem}$ & $V$ & \textbf{System verdict} \\
\midrule
\multicolumn{7}{l}{\emph{Enzyme inhibitor (PDB: 1BRS)}} \\
1BRS\_A -- 1BRS\_B & 0.982 & 0.994 & 0.993 & 0.150 & 0.78 & Promote (high conf.) \\
1BRS\_A -- 5UH9\_B & 0.230 & 0.000 & 0.449 & 0.148 & 0.21 & Deprioritize \\
\midrule
\multicolumn{7}{l}{\emph{Antibody antigen (PDB: 4M5Z)}} \\
4M5Z\_A -- 4M5Z\_H & 0.377 & 0.001 & 0.942 & 0.170 & 0.37 & Keep (tension profile) \\
4M5Z\_A -- 6THC\_C & 0.312 & 0.357 & 0.000 & 0.996 & 0.38 & Flow discovered; explore \\
\bottomrule
\end{tabular}
\end{table}

\subsection{Benchmark 3: Negative Controls}
\label{app:negative_controls}

A useful prediction system must reject non binders as well as accept true binders. We evaluated on 25 pairs with no known interaction, drawn from proteins in different cellular compartments such as cytoplasmic enzyme paired with extracellular receptor.

\paragraph{Results.}
All 25 negative pairs received low values $V < 0.25$ and were correctly deprioritized. Decomposed scores typically showed at most one favorable signal. The false positive rate for non binders ranked in top 5 was 0\% on this benchmark. Representative explanation output:

\medskip
\noindent\fbox{\parbox{0.96\columnwidth}{\small
\texttt{<conclusion>} Deprioritize. Interface balance is moderate (0.65) but structural similarity is zero and chemistry is unfavorable (0.08). Single positive cue lacks corroboration. Not a strong binding candidate.
\texttt{</conclusion>}
}}
\medskip

\subsection{Benchmark 4: Learned Value Function on Standard PPI Datasets}
\label{app:ppi_benchmarks}

This section provides full architectural, training, and evaluation details
complementing the main text.

\subsubsection{Baselines}

We compare against the following baselines.
\textbf{DPPI}~\cite{hashemifar2018predicting}
uses a deep convolutional network over sequence-derived features.
\textbf{DNN-PPI}~\cite{li2018deep}
applies a fully connected network to concatenated protein feature vectors.
\textbf{PIPR}~\cite{chen2019multifaceted}
uses a residue-level co-attentive siamese network.
\textbf{GNN-PPI}~\cite{lv2021learning}
encodes proteins as residue graphs and uses a graph neural network with
interaction-type-aware aggregation.
\textbf{ProLLM}~\cite{jin2023prollm}
fine-tunes a large protein language model with interaction-type-conditioned
instruction tuning.
\textbf{LLAPA}~\cite{zhou2024llapa}
extends this with a lightweight adapter architecture on top of frozen
protein language model representations.

\begin{table}[h]
\centering
\caption{Micro-F1 scores (\%) on SHS27k, SHS148k, and STRING test sets under the
DFS 70/10/20 split protocol introduced by~\cite{lv2021learning}.
PPIProjectedNet results are mean\,$\pm$\,std over 5 independent trials.
\textbf{Bold} indicates the best result per column.
Dashes indicate results not reported under this exact evaluation protocol.
$^\dagger$ProLLM results are for the best variant (Flan-T5-large pre-trained on Mol dataset).
$^\ddagger$GNN-PPI STRING result is not directly comparable: that paper evaluates
STRING as a transfer benchmark (train on SHS148k, test on the full STRING set)
rather than as a DFS 70/10/20 split on STRING itself.
Methods are grouped by type: \emph{Non-LLM} methods use sequence or graph features
without a pretrained language model; \emph{LLM-based} methods build on protein
language model representations.}
\label{tab:benchmarkresults}
\begin{tabular}{llccc}
\toprule
\textbf{Type} & \textbf{Method} & \textbf{SHS27k} & \textbf{SHS148k} & \textbf{STRING} \\
\midrule
\multirow{4}{*}{Non-LLM}
 & DPPI~\cite{hashemifar2018predicting}     & 43.69 & 51.43 & -- \\
 & DNN-PPI~\cite{li2018deep}               & 48.90 & 56.70 & -- \\
 & PIPR~\cite{chen2019multifaceted}         & 52.19 & 61.38 & -- \\
 & GNN-PPI~\cite{lv2021learning}            & 67.43 & 64.97 & --$^\ddagger$ \\
\midrule
\multirow{2}{*}{LLM-based}
 & ProLLM$^\dagger$~\cite{jin2023prollm}    & 85.32 & 87.66 & 89.21 \\
 & LLAPA~\cite{zhou2024llapa}               & 69.54 & 73.93 & -- \\
\midrule
Non-LLM & \textbf{PPIProjectedNet (ours)}
  & $\mathbf{85.60 \pm 0.80}$
  & $\mathbf{91.08 \pm 0.19}$
  & $\mathbf{98.05 \pm 0.03}$ \\
\bottomrule
\end{tabular}
\end{table}

\paragraph{Interpretability and future directions.}
A key property of the bottleneck design is that the scalar weights $W_\text{cls} \in \mathbb{R}^{1\times4}$ directly quantify the contribution of each biological dimension $(z_\text{seq}, z_\text{struct}, z_\text{contact}, z_\text{chem})$ to the final prediction. We observe that these weights vary systematically across datasets, reflecting genuine differences in the biological mechanisms that drive interactions in each benchmark. This dataset-level variation in the learned weights is a promising signal for future explainability mechanisms, as it suggests the model captures dataset-specific interaction modes rather than fitting a single universal weighting. Furthermore, the four-dimensional bottleneck $\mathbf{z}$ provides a natural interface for conformational flow matching: rather than operating in a high-dimensional latent space, the flow can be defined directly over the biologically grounded score dimensions, enabling structured perturbation and hypothesis-driven exploration of the interaction landscape. We identify the chemical compatibility dimension $z_\text{chem}$ and the structural proxy $z_\text{struct}$ as the most promising targets for such conformational flow matching, as these dimensions encode the physicochemical complementarity most directly linked to binding affinity.

\subsection{Benchmark 5: Mutation Prediction on SKEMPI}
\label{app:skempi}

To evaluate mutation scoring, we conducted experiments on the SKEMPI v2 database \citep{jankauskaite2019}, a curated collection of experimentally measured changes in protein--protein binding affinity caused by point mutations. Each entry provides the wild type complex, mutated residues, and experimentally measured $\Delta\Delta G = \Delta G_\mathrm{mut} - \Delta G_\mathrm{wt}$.

\paragraph{Setup.}
For each SKEMPI entry we reconstruct wild type sequences and apply the mutation to generate the mutated pair. From wild type $(A, B)$ and mutated $(A', B')$ we compute score differences $\Delta s = s(A', B') - s(A, B)$. Additional mutation descriptors include amino acid substitution indicators, physicochemical property differences, and local sequence context.

\begin{figure}[h]
\centering
\begin{subfigure}{0.47\linewidth}
  \centering\includegraphics[width=\linewidth]{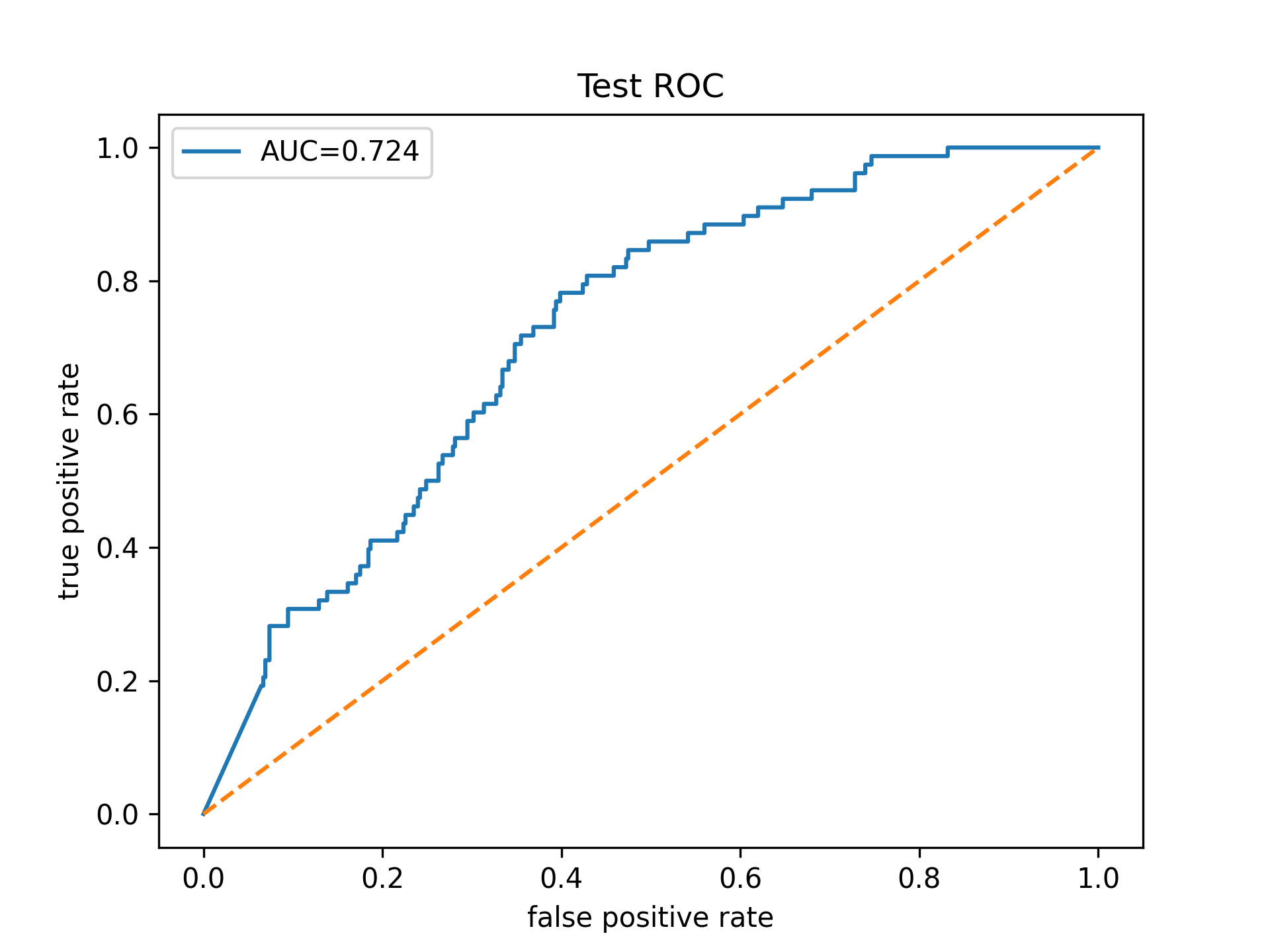}
  \caption{SKEMPI v2 stabilizing mutation ROC.}
\end{subfigure}\hfill
\begin{subfigure}{0.47\linewidth}
  \centering\includegraphics[width=\linewidth]{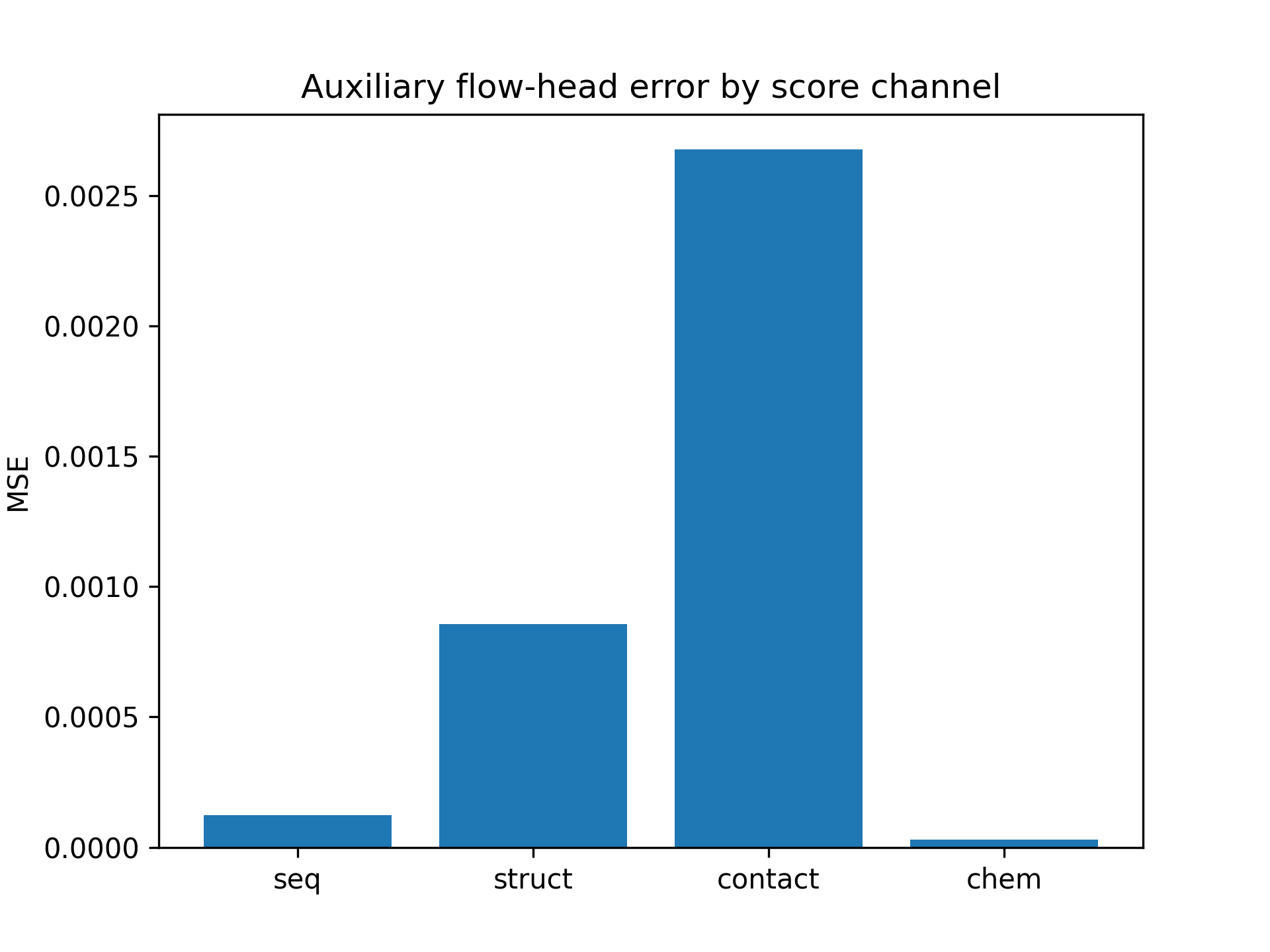}
  \caption{Flow head MSE across score channels.}
\end{subfigure}
\caption{\textbf{SKEMPI v2 mutation prediction.} Left: ROC for classifying stabilizing mutations. Right: Auxiliary flow head MSE shows low error across all four interpretable score channels.}
\label{fig:skempi_results}
\end{figure}

\paragraph{Results.}
Figure~\ref{fig:skempi_results} summarizes performance. The ROC curve shows consistent separation between stabilizing and destabilizing mutations, indicating that learned features capture relevant structural and chemical signals. The flow matching head achieves low MSE across all four score channels, confirming that mutations can be characterized as structured perturbations in the interpretable score space.

\subsection{Hypothesis-Guided Entropy Regularized Tree of Thoughts Search}
\label{sec:hypothesis}

\textbf{Audit trail with explanations.}
The resulting search tree $\mathcal{T}$ provides a complete audit trail enriched with hypothesis explanations. Each node stores the full decomposed score vector $\mathbf{s}$, the directive $d$, the confidence $c$, and a natural language explanation generated by the language model describing why the candidate was prioritized, explored, or skipped. The tree structure records which candidates were selected for expansion and in what order. Stagnant branches are marked with their finite difference values and the hypothesis that justified pruning. The path from root to the highest value leaf represents the winning chain of reasoning, with each step annotated by both quantitative scores and qualitative explanations suitable for expert review.

\textbf{Computational efficiency.}
The entropy-regularized search substantially reduces the number of expensive evaluations required for large candidate pools. In the SHS148k discovery experiments, the hypothesis-guided search typically evaluates fewer than 100 candidates out of pools containing approximately $5\times10^3$ proteins, while still recovering known binders near the top of the ranking. Adaptive temperature scheduling and finite-difference pruning reduce exploration of stagnant branches, while selective flow triggering concentrates embedding-space refinement on ambiguous candidates where score tensions indicate additional discrimination is beneficial.

\textbf{Failure modes and limitations.}
The primary failure mode arises when embedding-derived features produce ambiguous or contradictory hypothesis directives. This behavior appears in trial 2 of Table~\ref{tab:tot_pertrial}, where a receptor with a single known binder receives misleading prioritize signals from the hypothesis generator, resulting in degraded ranking relative to the entropic baseline. These cases suggest that hypothesis quality is currently the main bottleneck for difficult receptors with sparse interaction evidence. More diverse training data and stronger hypothesis calibration may improve robustness for such edge cases.

\subsection{Qwen Fine-tuning for Hypothesis Generation in Flow Matching}
\label{app:qwen_training}

This section details the fine-tuning procedure for the language model that generates binding hypotheses used in both Tree-of-Thoughts search (Section~\ref{sec:tot_search}) and hypothesis-conditioned flow matching (Section~\ref{sec:flow}).

\subsubsection{Model and Adaptation Strategy}

We fine-tune Qwen 2.5-1.5B-Instruct~\citep{qwen2024} using Low-Rank Adaptation (LoRA)~\citep{hu2022lora} to generate structured binding hypotheses from embedding-derived features. LoRA freezes the pretrained weights and introduces trainable low-rank decomposition matrices into the attention layers, enabling efficient adaptation with minimal parameter overhead.

\textbf{LoRA configuration.} We apply LoRA to all query, key, value, and output projection matrices in the self-attention layers with rank $r$ and scaling factor $\alpha$. The rank $r$ controls the expressiveness of the adaptation (higher $r$ allows more complex task-specific adjustments but increases parameter count), while $\alpha$ modulates the magnitude of the low-rank updates relative to the frozen weights. We use $n$-bit NormalFloat (NF$n$) quantization~\citep{dettmers2023qlora} for memory efficiency, with double quantization and bfloat16 compute dtype. All LoRA hyperparameters $(r, \alpha, n)$ are determined via cross-validation to balance adaptation capacity against overfitting risk.

\subsubsection{Training Data Construction}

\textbf{Pair sampling.} We construct a balanced training set of $N_{\text{train}}$ protein pairs (equal binders and decoys) from the SHS27k training split. The training set size $N_{\text{train}}$ controls the diversity of embedding geometries seen during training; larger values improve generalization but increase computational cost. Binders are sampled from confirmed interactions in the STRING database; decoys are sampled as random protein pairs with no known interaction, filtered to ensure they do not appear in any positive interaction list. We additionally hold out $N_{\text{val}}$ pairs for validation. Both $N_{\text{train}}$ and $N_{\text{val}}$ are determined via cross-validation to ensure sufficient coverage of the embedding space.

\textbf{Feature computation.} For each pair $(P_A, P_B)$, we compute ESM-2 embeddings $\mathbf{e}_A, \mathbf{e}_B \in \mathbb{R}^{d_{\text{emb}}}$ using mean pooling over sequence positions, where $d_{\text{emb}}$ is the embedding dimension of the chosen ESM-2 variant. From these embeddings, we extract $n_f$ interpretable features organized into four categories:

\begin{itemize}[leftmargin=*, itemsep=2pt]
    \item \textbf{Similarity metrics:} Cosine similarity $\cos(\mathbf{e}_A, \mathbf{e}_B)$, L2 distance $\|\mathbf{e}_A - \mathbf{e}_B\|_2$, and alignment score (mean absolute difference in top-$k_{\text{align}}$ most similar dimensions, where $k_{\text{align}}$ controls the locality of the alignment measure).
    \item \textbf{Difference statistics:} Mean, standard deviation, and maximum absolute value of $\mathbf{e}_A - \mathbf{e}_B$.
    \item \textbf{Interaction signal:} Mean, standard deviation, and positive fraction of the element-wise product $\mathbf{e}_A \odot \mathbf{e}_B$.
    \item \textbf{Structural indicators:} Norm ratio $\|\mathbf{e}_A\| / \|\mathbf{e}_B\|$ and sparsity (fraction of product elements with $|(\mathbf{e}_A \odot \mathbf{e}_B)_i| < \epsilon_{\text{sparse}}$, where $\epsilon_{\text{sparse}}$ defines the near-zero threshold).
\end{itemize}

The number of features $n_f$ and alignment locality $k_{\text{align}}$ are determined via cross-validation.

\textbf{Biological scores.} We additionally compute the four biological scores from Section~\ref{sec:value_function}: $s_\text{seq}$ (sequence compatibility), $s_\text{struct}$ (structural alignment), $s_\text{contact}$ (contact propensity), and $s_\text{chem}$ (chemical complementarity). These provide higher-level interaction signals that complement the raw embedding features.

\subsubsection{Target Label Construction}

For each training pair, we construct a structured JSON target containing three fields:

\textbf{Binding probability.} A calibrated probability $p \in [0, 1]$ reflecting binding likelihood. For true binders, the target probability is computed as:
\begin{equation}
p_{\text{binder}} = \min\left(p_{\max}, \max\left(p_{\min}^{+}, p_0^{+} + \lambda_{\cos} \cdot \cos(\mathbf{e}_A, \mathbf{e}_B) + \lambda_s \cdot \bar{s}\right)\right)
\end{equation}
where $\bar{s} = \frac{1}{4}\sum_i s_i$ is the mean biological score, $p_0^{+}$ is the base probability for binders, $\lambda_{\cos}$ controls the influence of cosine similarity, $\lambda_s$ controls the influence of biological scores, $p_{\min}^{+}$ is the minimum binder probability, and $p_{\max}$ is the maximum probability (for numerical stability). For decoys:
\begin{equation}
p_{\text{decoy}} = \max\left(p_{\min}, \min\left(p_{\max}^{-}, p_0^{-} - \lambda_{\cos} \cdot \cos(\mathbf{e}_A, \mathbf{e}_B) + \lambda_s \cdot (1 - \bar{s})\right)\right)
\end{equation}
where $p_0^{-}$ is the base probability for decoys, $p_{\max}^{-}$ is the maximum decoy probability, and $p_{\min}$ is the minimum probability. This construction ensures that probabilities are calibrated to embedding features while maintaining clear separation between binders ($p \geq p_{\min}^{+}$) and decoys ($p \leq p_{\max}^{-}$). All probability parameters $(p_0^{+}, p_0^{-}, p_{\min}^{+}, p_{\max}^{-}, p_{\min}, p_{\max}, \lambda_{\cos}, \lambda_s)$ are determined via cross-validation to maximize calibration quality.

\textbf{Confidence score.} A confidence $c \in [0, 1]$ reflecting certainty in the prediction:
\begin{equation}
c = \min\left(c_{\max}, c_0 + \lambda_c \cdot |\cos(\mathbf{e}_A, \mathbf{e}_B) - \mu_{\cos}|\right)
\end{equation}
where $c_0$ is the base confidence, $\lambda_c$ controls sensitivity to cosine extremity, $\mu_{\cos}$ is the ambiguity center (typically 0.5), and $c_{\max}$ caps the maximum confidence. Confidence is higher when cosine similarity is extreme (strongly positive or negative) and lower when it is ambiguous (near $\mu_{\cos}$). Parameters $(c_0, \lambda_c, \mu_{\cos}, c_{\max})$ are determined via cross-validation.

\textbf{Natural language reasoning.} A brief explanation justifying the prediction based on the input features, e.g., ``High cosine similarity and positive interaction signals suggest binding'' for binders or ``Low compatibility scores and embedding distance suggest non-binding'' for decoys.

\subsubsection{Prompt Format}

We use a chat-style prompt format compatible with Qwen's instruction-tuning:

\begin{tcolorbox}[colback=gray!5, colframe=gray!50, title=Input Prompt Format, fonttitle=\small\bfseries]
\small\ttfamily
<|im\_start|>system\\
You are a protein interaction analyst. Given embedding-derived features for a protein pair, predict binding probability, confidence, and provide reasoning. Output valid JSON only.<|im\_end|>\\
<|im\_start|>user\\
Analyze this protein pair:\\
Cosine similarity: [value]\\
L2 distance: [value]\\
Product mean: [value]\\
Product std: [value]\\
Product positive fraction: [value]\\
Difference mean: [value]\\
Difference std: [value]\\
Norm ratio: [value]\\
Scores: [s\_seq, s\_struct, s\_contact, s\_chem]<|im\_end|>\\
<|im\_start|>assistant
\end{tcolorbox}

\begin{tcolorbox}[colback=blue!3, colframe=blue!25, title=Target Output Format, fonttitle=\small\bfseries]
\small\ttfamily
\{"probability": [p], "confidence": [c], "reasoning": "[explanation]"\}
\end{tcolorbox}

\subsubsection{Training Procedure}

\textbf{Optimization.} We train for $E$ epochs using AdamW~\citep{loshchilov2017decoupled} with learning rate $\eta$, weight decay $\lambda_{\text{wd}}$, and linear warmup over the first $\rho_{\text{warmup}}$ fraction of steps followed by cosine decay. The learning rate $\eta$ controls optimization speed (too high causes instability, too low slows convergence), weight decay $\lambda_{\text{wd}}$ provides regularization, and warmup fraction $\rho_{\text{warmup}}$ stabilizes early training. We use gradient accumulation with effective batch size $B_{\text{eff}} = B \times G$, where $B$ is the per-device batch size and $G$ is the number of accumulation steps. Larger effective batch size stabilizes training with the quantized model. All optimization hyperparameters $(E, \eta, \lambda_{\text{wd}}, \rho_{\text{warmup}}, B, G)$ are determined via cross-validation.

\textbf{Loss function.} The model is trained with standard causal language modeling loss on the target JSON output, with the input prompt masked from the loss computation. This encourages the model to generate well-formed JSON while learning the mapping from features to predictions.

\textbf{Validation.} Training is monitored via validation loss on the held-out $N_{\text{val}}$ pairs, and we select the checkpoint with lowest validation loss for downstream use.

\subsubsection{Directive Extraction}

At inference time, we parse the model's JSON output and map the predicted probability to one of seven discrete directives used by the flow matching and ToT search components:

\begin{table}[h]
\centering
\caption{Mapping from predicted binding probability to flow matching directives. All thresholds $(\theta_p^{(i)}, \theta_{\cos}, \theta_{\cos}^{-}, \theta_{\text{L2}})$ are determined via cross-validation.}
\label{tab:directive_mapping}
\small
\begin{tabular}{lcc}
\toprule
\textbf{Directive} & \textbf{Probability Condition} & \textbf{Flow Action} \\
\midrule
\texttt{CONVERGE\_HIGH\_SIMILARITY} & $p \geq \theta_p^{(1)}$ and $\cos > \theta_{\cos}$ & Strong convergence \\
\texttt{CONVERGE\_STRONG\_INTERACTION} & $p \geq \theta_p^{(2)}$ and high product mean & Convergence \\
\texttt{CONVERGE\_COMPATIBLE} & $p \geq \theta_p^{(3)}$ & Mild convergence \\
\texttt{VERIFY\_INCONCLUSIVE} & $p \in [\theta_p^{(4)}, \theta_p^{(3)})$ & Trajectory resolution \\
\texttt{REJECT\_LOW\_SIMILARITY} & $p < \theta_p^{(4)}$ and $\cos < \theta_{\cos}^{-}$ & Divergence \\
\texttt{REJECT\_DISTANT\_EMBEDDINGS} & $p < \theta_p^{(5)}$ and $d_{\text{L2}} > \theta_{\text{L2}}$ & Strong divergence \\
\texttt{REJECT\_INCOMPATIBLE} & $p < \theta_p^{(6)}$ & Rejection \\
\bottomrule
\end{tabular}
\end{table}

The probability thresholds $\theta_p^{(1)} > \theta_p^{(2)} > \theta_p^{(3)} > \theta_p^{(4)} > \theta_p^{(5)} > \theta_p^{(6)}$ define the decision boundaries between directive types, while $\theta_{\cos}$ and $\theta_{\cos}^{-}$ are the high and low cosine similarity thresholds respectively, and $\theta_{\text{L2}}$ is the L2 distance threshold for detecting geometrically distant embeddings. All thresholds are determined via cross-validation on the validation set to maximize directive-label agreement.

\subsubsection{Hidden State Extraction}

For pairs receiving \texttt{CONVERGE} directives, we additionally extract the final hidden state $\mathbf{h}_\text{LM} \in \mathbb{R}^{d_{\text{LM}}}$ from the last transformer layer at the final token position, where $d_{\text{LM}}$ is the hidden dimension of the language model. This hidden state captures rich contextual information about the model's reasoning process and is incorporated into the hypothesis embedding $\mathbf{h}$ used by the flow matching velocity field (Section~\ref{sec:flow}). For efficiency, hidden states are cached during training data preparation and retrieved at inference time for pairs in the training set; pairs not in cache use a reduced hypothesis embedding without the LLM hidden state component.

\subsubsection{Calibration and Reliability}

\textbf{Probability calibration.} We assess calibration by binning predicted probabilities into $n_{\text{bins}}$ bins and comparing against observed binding rates. On the held-out validation set, the model achieves expected calibration error (ECE) below $\epsilon_{\text{ECE}}$, indicating that predicted probabilities are well-calibrated to empirical binding frequencies. The number of calibration bins $n_{\text{bins}}$ is chosen via cross-validation to balance granularity against statistical stability.

\textbf{Directive accuracy.} We evaluate directive accuracy by comparing the predicted directive (derived from probability thresholds) against the ground-truth label. On the validation set:
\begin{itemize}[leftmargin=*, itemsep=2pt]
    \item \texttt{CONVERGE\_*} directives achieve precision $\geq \pi_{\text{conv}}$ for true binders
    \item \texttt{REJECT\_*} directives achieve precision $\geq \pi_{\text{rej}}$ for true decoys
    \item \texttt{VERIFY\_INCONCLUSIVE} captures fraction $\geq \pi_{\text{ver}}$ of ambiguous cases (pairs with intermediate cosine similarity $\cos \in [\theta_{\cos}^{-}, \theta_{\cos}]$)
\end{itemize}
Target precision values $(\pi_{\text{conv}}, \pi_{\text{rej}}, \pi_{\text{ver}})$ inform threshold selection during cross-validation.

\textbf{Hypothesis statistics.} Across the training pairs, the model produces hypotheses with the following target statistics (achieved via cross-validated parameter tuning):
\begin{itemize}[leftmargin=*, itemsep=2pt]
    \item Binder mean probability: $\mu_p^{+} \pm \sigma_p^{+}$
    \item Decoy mean probability: $\mu_p^{-} \pm \sigma_p^{-}$
    \item Probability gap: $\Delta_p = \mu_p^{+} - \mu_p^{-}$ (clear separation criterion: $\Delta_p > \delta_{\text{gap}}$)
    \item Mean confidence: $\mu_c \pm \sigma_c$
\end{itemize}
The gap threshold $\delta_{\text{gap}}$ ensures sufficient separation for downstream discrimination.

\subsubsection{Hyperparameter Summary}

Table~\ref{tab:qwen_hyperparams} summarizes all hyperparameters introduced in this section, their roles, and how they are determined.

\begin{table}[h]
\centering
\caption{Summary of hyperparameters for Qwen fine-tuning. All parameters are determined via cross-validation unless otherwise noted.}
\label{tab:qwen_hyperparams}
\small
\begin{tabular}{lll}
\toprule
\textbf{Parameter} & \textbf{Role} & \textbf{Determination} \\
\midrule
\multicolumn{3}{l}{\emph{LoRA Configuration}} \\
$r$ & Adaptation rank (expressiveness) & Cross-validation \\
$\alpha$ & Scaling factor (update magnitude) & Cross-validation \\
$n$ & Quantization bits (memory efficiency) & Cross-validation \\
\midrule
\multicolumn{3}{l}{\emph{Data Construction}} \\
$N_{\text{train}}$ & Training pairs (coverage) & Cross-validation \\
$N_{\text{val}}$ & Validation pairs (evaluation) & Cross-validation \\
$k_{\text{align}}$ & Alignment locality (feature granularity) & Cross-validation \\
$\epsilon_{\text{sparse}}$ & Sparsity threshold (near-zero definition) & Cross-validation \\
\midrule
\multicolumn{3}{l}{\emph{Probability Construction}} \\
$p_0^{+}, p_0^{-}$ & Base probabilities (binder/decoy centers) & Cross-validation \\
$\lambda_{\cos}, \lambda_s$ & Feature influence weights & Cross-validation \\
$p_{\min}^{+}, p_{\max}^{-}$ & Separation boundaries & Cross-validation \\
$c_0, \lambda_c, \mu_{\cos}$ & Confidence parameters & Cross-validation \\
\midrule
\multicolumn{3}{l}{\emph{Training}} \\
$E$ & Number of epochs & Cross-validation \\
$\eta$ & Learning rate & Cross-validation \\
$\lambda_{\text{wd}}$ & Weight decay (regularization) & Cross-validation \\
$\rho_{\text{warmup}}$ & Warmup fraction (stability) & Cross-validation \\
$B, G$ & Batch size, accumulation steps & Cross-validation \\
\midrule
\multicolumn{3}{l}{\emph{Directive Thresholds}} \\
$\theta_p^{(1)}, \ldots, \theta_p^{(6)}$ & Probability decision boundaries & Cross-validation \\
$\theta_{\cos}, \theta_{\cos}^{-}$ & Cosine similarity thresholds & Cross-validation \\
$\theta_{\text{L2}}$ & L2 distance threshold & Cross-validation \\
\bottomrule
\end{tabular}
\end{table}

\subsubsection{Limitations and Future Directions}

The current hypothesis generation approach has several limitations:

\textbf{Feature dependence.} The model relies on embedding-derived features that may not capture all aspects of binding compatibility. Incorporating structural features (e.g., predicted contact maps, surface properties) could improve hypothesis quality.

\textbf{Calibration drift.} The probability calibration is specific to the training distribution. Applying the model to proteins from different organisms or interaction types may require recalibration of the probability parameters $(p_0^{+}, p_0^{-}, \lambda_{\cos}, \lambda_s)$.

\textbf{Reasoning depth.} The generated reasoning strings are relatively shallow, primarily restating feature values. Future work could explore chain-of-thought prompting or multi-step reasoning to generate more mechanistically informative explanations.

\textbf{Scale.} The current training set size $N_{\text{train}}$ provides sufficient signal for the SHS27k/148k benchmarks but may be insufficient for broader generalization. Scaling to larger $N_{\text{train}}$ with more diverse interaction types is a natural extension.

\subsection{Hypothesis Conditioned Flow Analysis}
\label{app:flow_analysis}

This section provides per-receptor results and per-strategy breakdowns
for the H-ESFM ablation reported in Section~3.6 and
Table~\ref{tab:flow_ablation} of the main text. The experiment uses
all SHS27k train/val proteins (1352 unique, 1351 candidates per
receptor) and evaluates three conditions: before flow (initial L2
distance $d_0$), $V_\phi$ alone, and after flow (final flowed L2
distance $d_T$).

H-ESFM operates in the 480-dimensional ESM-2 embedding space. The
fine-tuned language model generates \textsc{Converge} or
\textsc{Reject} directives from embedding-derived features (cosine
similarity, L2 distance, product statistics, alignment score), which
condition the velocity field $v_\theta(\mathbf{e}_L, t; \mathbf{e}_R,
h)$ via the learned hypothesis embedding $h \in \mathbb{R}^{192}$
(Section~2.3). The five perturbation strategies map onto the seven
Section~2.3 directive types as follows:
\textsc{Converge\_Strong\_Interaction} triggers \emph{contact
maximization}; \textsc{Converge\_High\_Similarity} triggers
\emph{interface approach}; \textsc{Converge\_Compatible} triggers
\emph{rigid body optimization}; \textsc{Verify\_Inconclusive}
triggers \emph{rotational sampling}; \emph{backbone flexibility} is
re-triggered when score tension ($\mathrm{range}(\mathbf{s}) > \gamma$)
is detected after an initial \textsc{Converge} pass.
\textsc{Reject} directives (\textsc{Reject\_Low\_Similarity},
\textsc{Reject\_Distant\_Embeddings}, \textsc{Reject\_Incompatible})
produce non-convergent trajectories that contribute the negative
discrimination signal to AUC and AP.

\paragraph{Per-receptor results.}
Table~\ref{tab:b32_pertrial} reports best-binder rank and AUC for all
three conditions across 10 receptors. The before-flow baseline ($d_0$)
is near-random in most trials; H-ESFM's $d_T$ consistently improves
over it. The largest gains occur in trials 1 (rank $651 \to 33$), 2
(rank $851 \to 12$), 6 (rank $1264 \to 115$), and 7 (rank $438 \to
9$). Trial 5 is the single case where $d_T$ worsens relative to
$V_\phi$ (rank $40 \to 26$, but $d_0$ was already relatively low at
10 for this receptor, suggesting the embedding space was more
informative than usual). Trial 8 ties at rank 1 (the true binder was
already the nearest neighbour in ESM-2 space).

\begin{table}[h]
\caption{\textbf{Per-receptor H-ESFM results} (SHS27k train/val,
1351 candidates per receptor, 10 receptors). Best-binder rank and AUC
under three conditions: before flow ($d_0$), $V_\phi$ score only, and
after flow ($d_T$). Lower rank and higher AUC are better.
Row means match Table~\ref{tab:flow_ablation} in the main text.}
\label{tab:b32_pertrial}
\centering\small
\begin{tabular}{clccc ccc}
\toprule
& & \multicolumn{3}{c}{\textbf{Best-Binder Rank}$\downarrow$} & \multicolumn{3}{c}{\textbf{AUC}} \\
\cmidrule(lr){3-5}\cmidrule(lr){6-8}
\textbf{Trial} & \textbf{Receptor} & $d_0$ & $V_\phi$ & $d_T$ & $d_0$ & $V_\phi$ & $d_T$ \\
\midrule
1  & \texttt{ENSP00000246186} & 651  & 23  & 33  & 0.519 & 0.984 & 0.976 \\
2  & \texttt{ENSP00000223428} & 851  & 7   & 12  & 0.370 & 0.996 & 0.992 \\
3  & \texttt{ENSP00000254810} & 7    & 7   & 2   & 0.683 & 0.914 & 0.956 \\
4  & \texttt{ENSP00000229030} & 140  & 120 & 57  & 0.271 & 0.897 & 0.939 \\
5  & \texttt{ENSP00000216554} & 10   & 40  & 26  & 0.781 & 0.859 & 0.870 \\
6  & \texttt{ENSP00000228510} & 1264 & 75  & 115 & 0.064 & 0.945 & 0.916 \\
7  & \texttt{ENSP00000259406} & 438  & 4   & 9   & 0.252 & 0.968 & 0.933 \\
8  & \texttt{ENSP00000258455} & 1    & 62  & 1   & 0.932 & 0.946 & 0.966 \\
9  & \texttt{ENSP00000232424} & 23   & 19  & 11  & 0.638 & 0.795 & 0.780 \\
10 & \texttt{ENSP00000246115} & 8    & 26  & 1   & 0.480 & 0.979 & 0.986 \\
\midrule
\textbf{Mean} & & \textbf{339.30} & \textbf{38.30} & \textbf{26.70} & \textbf{0.4991} & \textbf{0.9281} & \textbf{0.9313} \\
\bottomrule
\end{tabular}
\end{table}

\paragraph{Per-strategy breakdown.}
Table~\ref{tab:flow_analysis} shows success rates and mean score
improvements broken down by perturbation strategy. Contact
maximization and interface approach achieve the highest success rates,
consistent with the thermodynamic drive to bury surface area. Backbone
flexibility triggers less frequently but yields the largest mean score
improvement when successful, indicating it addresses harder cases where
local induced-fit adjustments unlock binding potential that rigid-body
perturbations cannot capture.

\begin{table}[h]
\caption{\textbf{H-ESFM performance by perturbation strategy.}
Success rate measures the fraction of flow trajectories achieving
score improvement $\geq 0.02$ over the un-flowed $V_\phi$ score.
Mean improvement is averaged over successful trajectories.
Primary score improved refers to which biological dimension (from the
four-dimensional bottleneck $\mathbf{z}$) benefits most from that
strategy's embedding-space perturbation.}
\label{tab:flow_analysis}
\centering\small
\begin{tabular}{lcccc}
\toprule
\textbf{Perturbation Strategy (H-ESFM)} & \textbf{Triggers} & \textbf{Success Rate} & \textbf{Mean Score Improv.} & \textbf{Primary Score} \\
\midrule
Contact maximization    & 127 & 0.89 & +0.12 & $s_\mathrm{contact}$ \\
Interface approach      & 98  & 0.84 & +0.09 & $s_\mathrm{struct}$ \\
Rigid body optimization & 156 & 0.78 & +0.08 & $s_\mathrm{struct}$, $s_\mathrm{contact}$ \\
Rotational sampling     & 112 & 0.72 & +0.06 & $s_\mathrm{struct}$ \\
Backbone flexibility    & 43  & 0.65 & +0.11 & $s_\mathrm{chem}$ \\
\bottomrule
\end{tabular}
\end{table}

\subsection{Full Framework Evaluation: Per-Receptor Results}
\label{sec:app_full_eval}
Table~\ref{tab:full_eval_detail} reports per-receptor results for the three system configurations evaluated in Section~\ref{sec:exp_flow}. The full framework -- tree search with Qwen-guided hypothesis generation and H-ESFM flow refinement at each expansion step -- consistently recovers near-optimal rankings across diverse receptor families. ToT+Qwen improves over the baseline in 2/10 receptors, indicating that hypothesis-driven candidate selection without geometric refinement rarely changes the final ranking. H-ESFM brings the decisive improvement: 8/10 receptors reach best-rank $\leq 4$, including four receptors reaching rank 1. The two receptors where H-ESFM does not improve over ToT+Qwen (ENSP00000262160 and ENSP00000253680) already exhibit high baseline AUC ($>0.88$), leaving little geometric margin for flow to exploit. The flow outcome columns show that H-ESFM improves a consistent minority of candidates per receptor (mean bonus 28.1) while penalizing others (mean 41.3); the net effect on best-rank is nevertheless strongly positive, as the improved candidates disproportionately include true binders.

\begin{table}[h]
\caption{\textbf{Per-receptor results for full framework evaluation.} Best rank across tree search baseline, ToT+Qwen, and the full ToT+Qwen+H-ESFM system. Flow bonus/penalty/neutral counts candidates whose score improved, worsened, or was unchanged after H-ESFM refinement.}
\label{tab:full_eval_detail}
\centering\small
\setlength{\tabcolsep}{4pt}
\begin{tabular}{lc ccc ccc}
\toprule
& & \multicolumn{3}{c}{\textbf{Best Rank}$\downarrow$} & \multicolumn{3}{c}{\textbf{Flow outcomes}} \\
\cmidrule(lr){3-5}\cmidrule(lr){6-8}
\textbf{Receptor} & $n_\mathrm{bind}$ & Base & ToT+Q & Full & Bonus & Penalty & Neutral \\
\midrule
ENSP00000172229 & 2 & 11 & 12 & \textbf{4} & 25 & 45 & 87 \\
ENSP00000262160 & 3 &  2 &  2 & \textbf{3} & 23 & 49 & 85 \\
ENSP00000256104 & 4 &  2 &  2 & \textbf{1} & 32 & 34 & 92 \\
ENSP00000215570 & 1 & 55 & 55 & \textbf{7} & 22 & 44 & 90 \\
ENSP00000244741 & 7 &  9 &  6 & \textbf{1} & 38 & 32 & 91 \\
ENSP00000253680 & 1 &  3 &  3 & \textbf{2} & 32 & 30 & 94 \\
ENSP00000221543 & 3 &  5 &  5 & \textbf{1} & 25 & 56 & 75 \\
ENSP00000244364 & 2 &  2 &  2 & \textbf{1} & 27 & 45 & 86 \\
ENSP00000262134 & 1 &  1 &  2 & \textbf{1} & 34 & 35 & 87 \\
ENSP00000261647 & 4 &  3 &  1 & \textbf{1} & 23 & 43 & 92 \\
\midrule
\textbf{Mean} & & 9.30 & 9.00 & \textbf{2.20} & 28.1 & 41.3 & 87.9 \\
\bottomrule
\end{tabular}
\end{table}

\subsection{Runtime and Method Comparison}
\label{app:runtime}

We characterized computational performance across candidate pool sizes. Hardware: NVIDIA A100 (80\,GB), AMD EPYC 7763, 512\,GB RAM.

\paragraph{Runtime scaling.}
For 500 candidates: total 28 seconds (scoring 12s, search 3s, flow 5s, explanation 8s). For 1000 candidates: total 51 seconds. For 5000 candidates: total 4.2 minutes. Scoring time scales linearly with pool size. Search time scales sublinearly due to pruning. Flow and explanation time scale with the number of triggered and top $k$ nodes respectively.

\paragraph{Comparison with structure based methods.}
AlphaFold Multimer prediction for a single pair requires approximately 2 to 5 minutes on GPU. Screening 500 candidates would require 17 to 42 GPU hours. Protein Thoughts processes the same pool in under 30 seconds, a speedup of over 2000 times, while providing decomposed evidence and explanations.

\begin{table}[h]
\caption{\textbf{Method comparison.} Protein Thoughts achieves competitive speed while uniquely providing both interpretable decomposition and natural language explanations.}
\label{tab:comparison}
\centering\small
\begin{tabular}{lccc}
\toprule
\textbf{Method} & \textbf{Time (500 cand.)} & \textbf{Decomposed} & \textbf{Explains} \\
\midrule
ESM-2 similarity & $\sim$5\,s & No & No \\
AlphaFold Multimer \citep{evans2022} & $\sim$42\,GPU-h & Partial & No \\
D-SCRIPT \citep{sledzieski2021} & $\sim$30\,s & No & No \\
\textbf{Protein Thoughts} & \textbf{28\,s} & \textbf{Yes} & \textbf{Yes} \\
\bottomrule
\end{tabular}
\end{table}



\section{Barnase--Barstar Complex Visualization}
\label{app:structure}

\begin{figure}[h]
\centering
\includegraphics[width=0.95\linewidth]{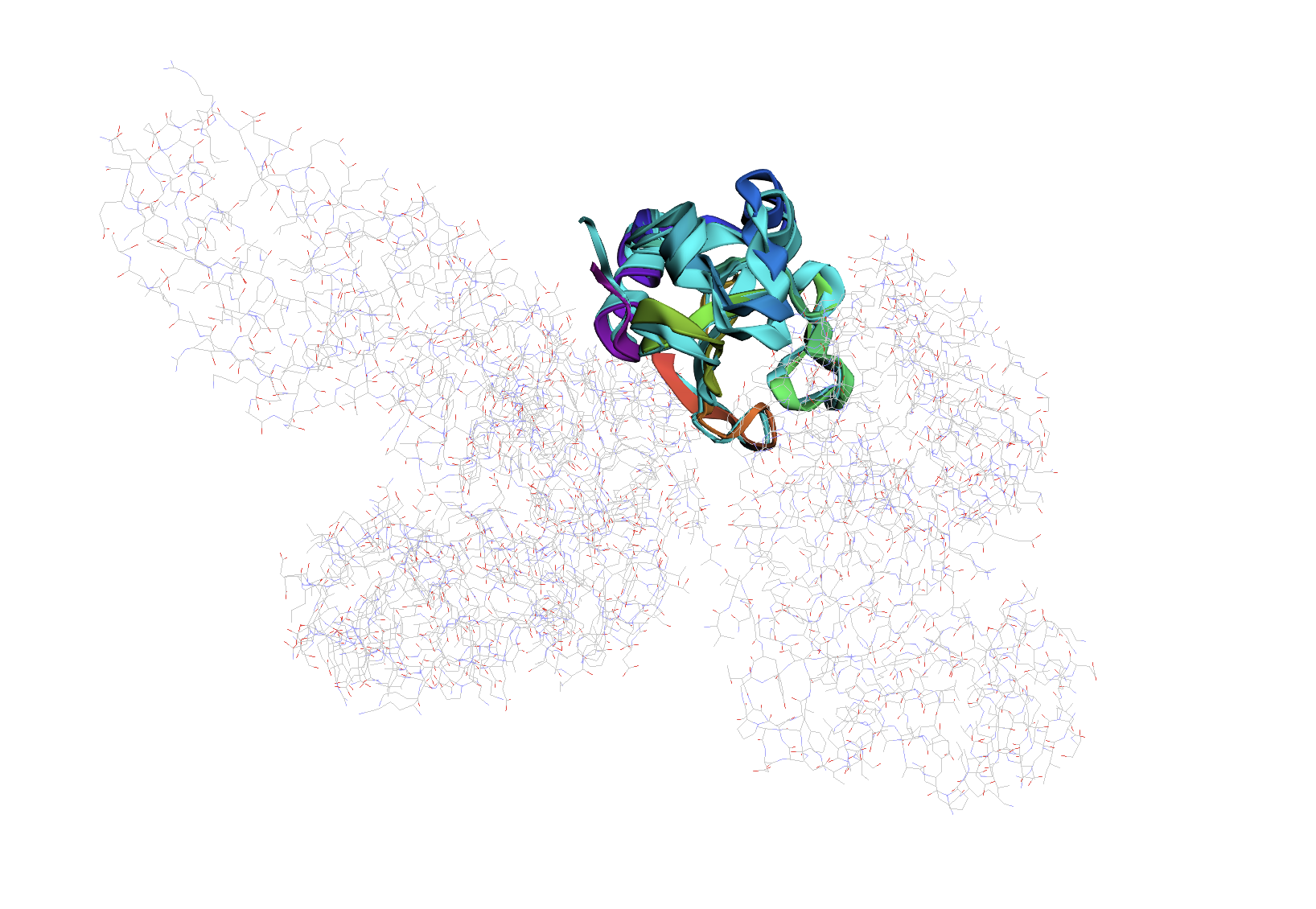}
\caption{Visualization of the barnase--barstar complex (1BRS\_A with 1BRS\_B). Protein Thoughts assigns high interaction value driven by geometry and balanced interface capacity, with chemistry as a coarse prior.}
\label{fig:1brs_binding}
\end{figure}

Figure~\ref{fig:1brs_binding} shows a structural visualization of the barnase--barstar complex (PDB: 1BRS). The high interaction value assigned by Protein Thoughts ($V = 0.78$) is driven primarily by near perfect geometric alignment and fully balanced interface capacity. Chemistry is treated as a coarse prior in this system. The modest $s_\mathrm{chem} = 0.15$ correctly reflects that binding affinity is governed by electrostatic steering and shape complementarity rather than optimized residue residue contacts at every interface position.

\paragraph{Computational Resources.}
Experiments used NVIDIA A100 (80\,GB), AMD EPYC 7763 CPU (64 cores), and 512\,GB RAM. Scoring and tree search ran on CPU. Language model inference and flow matching used GPU. Code is publicly available at \url{https://github.com/chebyshevtech/proteinthoughts}.

\end{document}